\documentclass[prd,aps,amssymb]{revtex4}

\usepackage{epsfig,amsmath,amsfonts}
\usepackage{dsfont}
\usepackage[cp1251]{inputenc}
\usepackage[english]{babel}
\usepackage{epstopdf}
\usepackage[]{graphicx}
\usepackage{subfigure}
\newcommand{\be}{\begin{equation}}
\newcommand{\ee}{\end{equation}}
\newcommand{\bn}{\begin{eqnarray}}
\newcommand{\en}{\end{eqnarray}}
\newcommand{\bd}{\begin{displaymath}}
\newcommand{\ed}{\end{displaymath}}

\begin{document}

\title{Modified equations in the theory of induced gravity. Solution to the cosmological constant problem }
\author{Farkhat Zaripov}
\email{farhat.zaripov@kpfu.ru}
\affiliation{Institute of Mathematics and Mechanics, Kazan Federal University, 18 Kremlyovskaya St., \\ Kazan, 420008, Russia\\ farhat.zaripov@kpfu.ru}

\begin{abstract}
This research is an extension of the author's works , in which conformally invariant generalization of string theory was suggested to higher-dimensional objects. Special cases of the proposed theory are Einstein's theory of gravity and string theory.

 This work is devoted to the formation of self-consistent equations of the theory of induced gravity  in the presence of matter in the form of a perfect fluid that interacts with scalar fields. The study is done to solve these equations for the case of the cosmological model.
 In this model time-evolving gravitational and cosmological ''constants'' take place which are determined by the  square of scalar fields. The values of which can be matched with the observational data.

 The equations that describe the theory have solutions that can both match with the solutions of the standard theory of gravity as well as it can differ from it. This is due to the fact that the fundamental "constants'' of the theory, such as gravitational and cosmological, can evolve over time and also depend of the coordinates.
Thus, in a rather general case the theory describes the two systems (stages): Einstein and "evolving''. This process is similar to the phenomenon of phase transition, where the different phases (Einstein gravity system, but with different constants) transit into each other.

\end{abstract}
\maketitle

\section{Original field theory}

There are arguments \cite{brn} in favor of the theory of physical fields that it must have the property of conformal invariance, at least at the classical level, up to the time when this symmetry is not broken. The action for the membrane ($n-1$-branes) admits no conformal transformations and for these models there is no natural candidate for the role of anomalous symmetry rather than a conformal symmetry for the string theory. To avoid this difficulty while remaining within the ideology of the string theory we proposed the following generalization of string theory \cite{zar}, \cite{zari}
\begin{equation}\label{ns1}
 \frac{1}{w} \int \left\{ - \frac{1}{2}
(\nabla_{\nu}{X},\nabla^{\nu}{X})+{\xi}{R}(X,X)+U \right\}
 \sqrt{- g} \hat d^{n} \sigma.
\end{equation}
The following notation is adopted:

$(X,X)=$ $X^{A}X^{B}\eta_{AB}$ $\equiv Y,$ \quad
$ (\nabla_{\nu}{X},\nabla^{\nu}{X})=
\nabla_{\nu}X^{A}\nabla_{\mu}X^{B}g^{\nu\mu}\eta_{AB}, $ \quad
$U=U(X^{A})$ - fields-dependent potential $X^{A}$. In the article \cite{zar} we used
$U(X^{A})=U_{0}\equiv \Lambda_X (X,X)^{\rho}$, where $\rho=\frac{n}{n-2}$.
For simplicity let's assume $U(X^{A})=U(Y(X^{A}))$ .
For action (\ref{ns1}), the functions
$X^{A}=X^{A}(\sigma^{\mu})$, where $A,B=1,2, \ldots , D,\quad
\mu,\nu=0,1,\ldots,n-1$, map $n$-dimensional manifold $\Pi$
described by the metric $g_{\mu\nu},$ into $D$-dimensional space-time $M$ with the metric $\eta_{AB},$ where space $M$ described by the Minkovski metric with the signature $(-,+,\ldots,+)$. However, as it turns out on closer examination it is more convenient to leave $M$ signature of an arbitrary. The flat space signature is understood here as the set of signs of the elements along the main diagonal ($+1$ and $-1$) of the metric matrix. $R$ is the scalar curvature of the manifold $\Pi$, the operator $\nabla_{\nu}$ denotes the covariant derivative in the manifold $\Pi$,
where the Christoffel symbols connected with the metric in the standard way. Let's assume that the space of $\Pi$ is parameterized by the coordinates $\sigma^{\mu}$, where $\sigma^{0}= t$ is the time coordinate, and the components $\sigma^i$ $(i=1,2,\ldots,n-1)$ describe a certain $n-1$ - dimensional object.
The values of $w$, ${\tilde\xi}$, $\Lambda_X$ are constant.
Action (\ref{ns1}) has the property of conformal invariance if
\begin{equation} \label{ns2}
{\xi}=-\frac{n-2}{8(n-1)}
\end{equation}
(when $n=4$, ${\xi}=-\frac{1}{12}$).
 This invariance is expressed in the fact that the equations obtained by
  varying the action (\ref{ns1}) with respect to the fields $\hat g$ and
  $\hat X$ are invariant under the local Weyl scale changes
\begin{equation} \label{cp}
g_{\mu\nu} \Longrightarrow e^{2\phi} g_{\mu\nu}, \qquad X^{A}
\Longrightarrow e^{4\xi (n-1)\phi}X^{A},
\end{equation}
for an arbitrary function $\phi = \phi(\sigma^{\mu})$.

After varying the action (\ref{ns1}), the field equations for
  $\hat X$ and $\hat g$ have the following form::
\be \label{Y}
 \Box X^{A} + 2 \xi R X^{A} + 2 \frac{d U}{d Y} X^{A} = 0,
\ee
\bd
 T_{\alpha\beta} \equiv  \frac{1}{w} [(\nabla_{\alpha}X,
\nabla_{\beta}X) - \frac{1}{2}(\nabla_{\mu} X, \nabla^{\mu}
X)g_{\alpha \beta} + U g_{\alpha\beta}]+
\ed
\be \label{T}
+\frac{2\xi}{w}[ -
 R _{\alpha\beta} + \frac{1}{2} R g_{\alpha
\beta} + \nabla_{\alpha} \nabla_{\beta} - g_{\alpha\beta} \Box ]
(XX) = 0,
\ee

If the action is supplemented by Lagrange functions of other matter
  fields, then (\ref{T}) is replaced by the equation
\be \label{T3}
T_{(tot)\alpha\beta} \equiv T_{\alpha\beta} + T_{(e)\alpha\beta}=0
\ee
where $T_{\alpha\beta}$ and $T_{(e)\alpha\beta}$ - are energy-momentum tensors (EMT) of fields $X^{A}$ and other fields of matter (such as a perfect fluid), respectively.

Overview of various  F (R)  theories including the theory of induced gravity are given in \cite{odints2}, \cite{odints3}.

If $(X,X)=const$, equations (\ref{T3}) and (\ref{T}) are similar to the Einstein equations with an effective gravitational constant
 \be \label{kap}
 k_{eff} = \pm\frac {w c^{3}} {16\pi\xi (X, X)\hbar}\equiv G_{eff}\frac{c^3}{8 \pi \hbar},
  \ee
  $\hbar$ - Planck constant. Let's compare the experimental value of gravitational constant $G_0$ with the effective "gravitational constant".
\be \label{Eg4}
G_0\equiv\frac{8 \pi k_n \hbar }{c^3}\equiv6.565362\cdot 10^{-65}cm^2=\pm\frac {w } {2\xi Y_{mod} },
\ee
where $Y_{mod}$ - modern cosmological value of function  $Y;$  $k_n=6.674286 \cdot 10^{-8} cm^3 c^{-2} g^{-1}$ - Newton's gravitational constant.
Henceforth we shall use the system of units: $c=1,  \quad \hbar=1.$
To find out the sign in $G_{eff}$ requires further research.
If we consider generalization of the Einstein's equation with the standard one-component scalar field \cite{landau}, it is necessary to choose a sign $+$, then $\xi
(X, X)>0.$ However this sign is dependent of the signature of space-time $M$. This article considers the various choices of that sign. In general, we can assume that the sign of the parameter $w$ of an arbitrary.

We note the important fact that for strings ($n = 2$) the general solution of (\ref{T}) is:
\be \label{gtti}
B_0g_{\mu \nu} = (\nabla_{\mu} X, \nabla_{\nu} X) \quad \mu, \nu =
\overline{0,n-1},
\ee
where $B_0$ of an arbitrary function.
Therefore, the metric $g _ {\mu \nu}$ of manifold $\Pi$ is connected by a conformal transformation with the induced metric $ (\nabla_{\mu} X, \nabla_{\nu} X) $ on the surface $X^A=X^A (\sigma^i)$ $(i=0,1,\ldots, n-1)$. When $B_0=const\neq 0 \quad(B_0=1)$ equations (\ref{gtti}) are the conditions of embedding the manifold $\Pi$ into multidimensional flat space-time $M$. Unfortunately in general case $n > 2$ for (\ref{Y}) - (\ref{T}), the solution (\ref{gtti}) is not a general solution. The problem of connection between the metric of the manifold $\Pi_g$ with the metric induced by the solutions $X^A = X^A (\sigma^{\mu})$, as well as that of a physical interpretation  of this connection, have not been solved for an arbitrary dimension. However, in the case of dimension $n=4$, as follows from the analysis of Hamilton's equations, the solution of (\ref{gtti}) can be regarded as gauge fixing. In the cases of additional symmetries associated with the cosmological solutions we can fix the value $B_0=1$.
 In the article \cite{zar} we considered some particular cosmological solutions for equations
 (\ref {Y}) - (\ref {T3}), (\ref {gtti}).

\section{Macroscopic equations}
Note the following statement: for conformally invariant case
(\ref{ns2}), considering equation (\ref{Y}), EMT trace $\quad T_{\alpha\beta}$ is equal to zero.  As follows from equation (\ref{T3}), the trace $T_{(e)\alpha\beta}$ is also equal to zero - meaning that there can exist matter only of ultra-relativistic equation of state $\varepsilon =3p$. To consider matter with the different equations of state it is necessary to modify the theory. We shall proceed from two assumptions.

 1)Conformal invariance maybe broken due to the evolving behavior of the interaction parameter $\xi$. One can recall the theory of the renormalization group \cite{odints1} in the theory of quantization...

2)Conformal invariance is clearly not broken, but through interaction of fields $X^{A}$ with matter fields in the equations appear additional terms. Under the notion of "material fields" we assume spinor fields and vector fields, the latter are the carriers of interaction. We shall also assume that the averaged form of EMT of these fields and their interactions has a structure of EMT of ideal fluid.
\be \label{ep}
 T_{(e)\alpha}^{\beta}=(p+\varepsilon)u_{\alpha}u^{\beta}+p\delta_{\alpha}^{\beta},
 \ee
 where  $u_{\alpha}$ $4$ - speed and $u_{\alpha}u^{\alpha}=-1$.

 Due to interaction with vector fields equation (\ref {Y}) acquires additional term $S^{A}$. So the equation takes form
\be \label{Y1}
 \Box X^{A} + 2 \xi R X^{A} + 2 \frac{dU}{dY} X^{A}  = S^{A}.
\ee In this article we will not clarify the specific form of this additional term, but consider in a general way.It is assumed that the contribution from terms $S^{A}$ in EMT contained in $T_{(e)\alpha}^{\beta}.$ The equation of state of matter $\varepsilon=\varepsilon(p)$ is determined by additional conditions. With all this let's assume that a general law of conservation takes its place:
\be \label{T4}
 \nabla_{\beta}T_{(tot)\alpha\beta} \equiv \nabla_{\beta}(T_{\alpha}^{\beta} + T_{(e)\alpha}^{\beta})=0.
  \ee
  Thus, considering the system of equations (\ref {gtti}), (\ref{T3}), (\ref{Y1}), (\ref{T4}). Let's define the function $Y=(X,X)$.
 From equation (\ref{Y1}) by scalar multiplication by $X_{A}$ we get:
\be \label{Y2}
 \Box Y -2nB_0 + 4 \xi R Y + 4 \frac{d U}{d Y} Y  = 2(S,X).
\ee
The EMT, given the previous equation, can be rewritten as:
\bd
w T_{\alpha}^{\beta}=-2\xi Y[ R _{\alpha}^{\beta} - \frac{1}{n}R
\delta_{\alpha}^{\beta}]+2\xi[\nabla_{\alpha} \nabla^{\beta}Y -
\frac{1}{n} \delta_{\alpha}^{\beta} \Box Y]+
\ed
\be \label{T5}
+\frac{1}{2n}\delta_{\xi}\delta_{\alpha}^{\beta}\Box Y +
\frac{n-2}{2n}(S,X)\delta_{\alpha}^{\beta}+(U-\frac{n-2}{n}Y\frac{d U}{d Y} )\delta_{\alpha}^{\beta}.
\ee
Here we introduce a new parameter $\delta_{\xi}$:
$$\xi=-\frac{n-2}{8(n-1)}-\frac{\delta_{\xi}}{4(n-1)}\,$$ which characterizes the deviation from the conformally invariant case. For the dimension $n=4$:
$\delta_{\xi}=-12 \xi -1$.

The law of conservation (\ref{T4}) takes the form:
\be \label{T6}
 (S,\nabla_{\beta}X )+ w\nabla_{\beta}T_{(e)\alpha}^{\beta}=0.
  \ee
From equation (\ref{Y2}) and differential consequences of condition
(\ref {gtti}) we get an equation:
\be \label{Y3}
 \nabla_{\beta}Y \cdot (\xi R+\frac{d U}{d Y})=
 (S,\nabla_{\beta}X )+\frac{n-2}{2} \nabla_{\beta}B_0.
  \ee
 Taking into account the last equation, the law of conservation can be rewritten as:

\be \label{T7}
 -\frac{n-2}{2} \nabla_{\beta}B_0 + \nabla_{\beta}Y \cdot
 (\xi R+\frac{dU}{dY})+w\nabla_{\alpha}T_{(e)\beta}^{\alpha}=0.
  \ee
Trace of the total EMT equals zero:
\be \label{T8}
w T_{(tot)\alpha}^{\alpha} \equiv  \delta_{\xi}[B_0 n -2 \xi R Y] -4\xi(n-1)(S,X)+Un+8(n-1)\xi Y\frac{dU}{dY}+wT_{(e)\alpha}^{\alpha}=0
\ee
Excluding term $(S,X)$ from equations (\ref{Y2}) and (\ref{T8})and
including it into equations (\ref{Y2}) and (\ref{T5}) we get:
\be \label{Y4}
\Box Y = \frac{n-2}{4(n-1)\xi}[-nB_0 + 2 \xi R Y +
\frac{2n}{n-2}U]+\frac{w}{2\xi(n-1)}T_{(e)\alpha}^{\alpha}
\ee

\be \label{E}
G_{\alpha \beta}=\frac{1}{2\xi Y}[-\frac{n-2}{2}B_0 +U]g_{\alpha \beta}+\frac{1}{Y}[\nabla_{\alpha}\nabla_{\beta}- g_{\alpha\beta} \Box]Y+ \frac{w}{2 \xi Y}  T_{(e)\alpha\beta},
\ee
where $G_{\alpha \beta}$ - the Einstein tensor.
Equation (\ref{E}) - is the analogue of Einstein equations for the macroscopic environment. Function $Y$ is a solution of equation (\ref{Y4}), however, this equation is an algebraic consequence of the equations (\ref{E}). Differential consequence of these equations
is the law of conservation - which has the form (\ref{T7}).

Equations (\ref{E}), except for the first term on the right side of the equation, are the same as a special case of Brans-Dicke. If a scalar field in this theory is:
$\Phi=16 \pi \xi Y/w$. Fields $X^{A}$ as coordinates of space $M$ have the dimension in centimeters, from which it follows $[Y]=cm^{2}$ and $[w]=cm^{4}$. Action (\ref{ns1}) we took as a dimensionless quantity.

The first term in the equations (\ref{E}) can be interpreted as a "cosmological constant". However, to take into account the effect of the energy of polarization of vacuum into gravity, we need to highlight from EMT matter a part related to this energy, which satisfies the state of equation: $\varepsilon_{vac}+p_v=0$, where $\varepsilon_{vac}$ and $p_v$ - are energy density and pressure of the vacuum polarization. It is assumed that the vacuum (on average) has the properties of homogeneity and isotropy. The density of vacuum energy related with quantum effects is associated (in QFT) with ultraviolet cutoff. Lets assume that UV cutoff is carried out at the Planck scale \cite{vain}
\be \label{evac}
\varepsilon_{vac}=(\frac {8 \pi }{G_{0} })^2\simeq 1.5 \cdot 10^{131}cm^{-4} \Rightarrow
\ee
$\Lambda_{vac}= G_{0} \varepsilon_{vac}= 9.62 \cdot 10^{66}cm^{-2}.$
Then in the equations (\ref{E}) term related to the effective "cosmological constant" takes form:
 \be \label{lamcos1}
\Lambda_{eff}= \frac{1}{2\xi Y} (-\frac{n-2}{2}B_0+w \varepsilon_{vac}+U).
 \ee
 In future we use the notation $B=\frac{n-2}{2}B_0-w \varepsilon_v$ and in the equations (\ref{T7}),(\ref{Y4}), (\ref{E}) we need to make substitution $B_0\Rightarrow B $.

Thus, there are evolving over time gravitational ($G_{eff}$) and  cosmological  ($\Lambda_{eff}$)  ''constants'':
\be \label{E4}
G_{eff}= \frac {w } {2\xi Y },\quad \quad \Lambda_{eff}
= \frac{1}{2\xi Y} (-B+U),\quad n=4,
\ee

In general, we get the systems of "macroscopic" equations (\ref{E}), ``microscopic'' equations (\ref{Y1}) and constraint equations (\ref{gtti}), where $Y=(X,X)$. The study of the complete system requires the definition of the model, ie definition of $S^{A}$. In this article, we will not deal with this issue, noting only naturally occurring hypothesis that specific sector of fields $\{X^1,X^2,...X^k\},\quad k<D$
can play a role of Higgs scalar field theory of strong and electroweak interactions. We note that the gravitational equations (\ref{E}) include only macroscopic features: $Y, B, g_{\alpha\beta}$ and EMT material environment $T_{(e)\alpha\beta}$.

\begin{em}
Remarks:

 1. The resulting equations (\ref{T7}),(\ref{Y4}), (\ref{E}) are valid for more general case, where $B$ is a function of coordinates of the manifold.

 2. The condition of "embedding" (\ref{T7}), which was used in deriving equations, does not limit the proposed theory. Indeed, in the most general case, we can make the substitution:
 \be \label{gtti1}
(\nabla_{\mu} X, \nabla_{\nu} X)=B_0 g_{\mu \nu}+k_{\mu \nu}  \quad \mu, \nu =
\overline{0,n-1},
\ee
where  $k_{\mu \nu}$ - some tensor functions. If you do the calculations described above, it can be shown that the resulting equations have the same form as in equations (\ref{T7}),(\ref{Y4}), (\ref{E}), if in these equations redefine EMT matter
\be \label{gtti2}
T_{(e)\alpha\beta} \Rightarrow T_{(e)\alpha\beta} +\frac{1}{w}(k_{\alpha \beta} -\frac{1}{2} g_{\alpha\beta} k_{\mu \nu} g^{\mu \nu} ).
\ee

 3. Equation (\ref{Y4}) is an algebraic consequence of the equations (\ref{E}), and the equation (\ref{T7}) is the differential consequence of equations (\ref{E}). Thus, the number of unknowns is greater than the number of independent equations. However, additionally is given equation of state of matter and the structure of equations of "the law of conservation" (\ref{T7}) such that provides additional equations after the imposition of certain conditions related to the equation of state.
 \end{em}
In the case of "embedding" $(B_0=1)$,  as follows from equations (\ref{T7}),  following cases are possible:

 I)$Y=C=const,$  $\nabla_{\beta}T_{(e)\alpha}^{\beta}=0.$

Note also that $Y=C=const, B=const.$  $\Rightarrow$ $\nabla_{\beta}T_{(e)\alpha}^{\beta}=0.$

In this case, we obtain the equations coinciding with the equations of Einstein, gravitational constant $G_{eff}=const$ and the cosmological constant $\Lambda_{eff}=const$.

Equations (\ref{Y1}) can be rewritten as:

\be \label{CY1}
 \Box X^{A} -( -4 \frac{B}{C} +\frac{w}{C}(\varepsilon
-3P)) X^{A}  = S^{A}
\ee - free fields ($S^{A}=0$) $X^A$
acquire mass $\mu$, where $\mu^2=-4\frac{B}{C}
+\frac{w}{C}(\varepsilon -3P)$.

 II) $Y \neq const,$ \emph{and separate law of conservation of matter is performed}:
 $\nabla_{\beta}T_{(e)\alpha}^{\beta}=0.$  In this case from
(\ref{T7}) follows the equation
\be  \label{sl2}
 \xi R+\frac{d U}{d Y}=0.
\ee
Equations (\ref{Y1}) can be rewritten as:
\be \label{CY1Y}
 \Box X^{A} = S^{A}\ee - free fields ($S^{A}=0$) $X^A$ have zero mass.

There is also a third case:

III) When $Y \neq const, \quad B=const$ \emph{and separate law of conservation of matter not necessarily fulfilled}.  This case is a generalization of the previous case. The law of conservation takes the form:
\be \label{T72}
 - \nabla_{\beta}Y \cdot
 (\xi R+\frac{d U}{d Y})=w\nabla_{\beta}T_{(e)\alpha}^{\beta}.
\ee

\subsection{Cosmological solutions}
Let's consider metric form of the manifold $\Pi$ corresponding to homogeneous and isotropic cosmological model
 \be  \label{met} d s^2 = -a^2 (\eta) [ d \eta^2 - (d
\chi)^2 - K (\chi)d \Omega^2],
\ee
where   $K (\chi) =\{ \sinh^2
{\chi};\sin^2 {\chi};\chi^2\}$ - respectively, for the models of open, closed and flat types. $d \Omega^2$ - metric form of unit sphere, expressed in spherical coordinates.

\subsubsection{vacuum solutions}

Let's consider the above equations under potential
$$U =\Lambda (X,X)^{2} \equiv \Lambda_X Y^{2} ,$$
($n=4;\quad \rho=2; \quad B_0=1.$)

I) $Y=C=const.$
 In the case of vacuum ($S^A=0, \quad T_{(e)\alpha\beta}=0$) these equations can be analytically solved (\ref{Y4}), (\ref{E}), (\ref{gtti}).

 For the fields $X^{A}$ we get the following equations
\be \label{tah}
 \Box X^{A} + 4\frac{B}{C} X^{A}= 0.
\ee
Found (\cite{zar}) particular solutions of these equations satisfying the conditions of "immersion" (\ref{gtti}). In a closed model these solutions have the form:
\be \label{massht}
 a(t)=\frac{\cosh(t H)}{H},
 \ee
where $H^2=(B/C),$   $t$  is proper time  ($dt=a(\eta)d\eta$).
\be \label{zakritx}
X^{0} = \sqrt{C} \tan {(\eta+\eta_0)}, \quad X^a =
\frac{\sqrt{C}}{\cos {(\eta+\eta_0)}} k^a,
\ee
 where  $k^a $- immersion function of  $3$ - dimension sphere
 \bd
 k^1 = \sin {\chi} \sin {\theta} \cos {\phi},\qquad k^2 =
\sin {\chi} \sin {\theta} \sin {\phi},
\ed
\be \label{ka} k^3 = \sin
{\chi} \cos {\theta}, \qquad k^4 = \cos {\chi}.
\ee
Conditions  (\ref{gtti})  also determine the relationship between constants:
  \be \label{coslam} \Lambda_X = \frac {B(3+\delta_{\xi})}{2C^2},  \quad
\Lambda_{eff} = \frac {3 B}{C}.
\ee
 Note the important fact that $\Lambda_{eff}$ is independent from $\xi$ and $\delta_{\xi}=-3$ $\Rightarrow \Lambda_X=0 $ .

For the closed model manifold $\Pi$ forms a "hyperboloid of one sheet" in the space of $M$, and is defined by the equation:
\be \label{constc}
-(X^0)^2+(X^1)^2+(X^2)^2+(X^3)^2+(X^4)^2=C.\ee
 For the case of an open model there are two possibilities: a) $C/B<0$ space immersed in 5-dimensional flat space with signature $(-,+,+,+,-)$ and the scale factor has the form
 \be \label{masshto1}
 a(t)=\frac{\cos(t |H|)}{|H|},
 \ee
 b)$C/B>0$ - signature of flat space -$(+,+,+,+,-)$, and the scale factor has the form
  \be \label{masshto2}
 a(t)=\frac{\sinh(t H)}{H}.
 \ee
 Solutions for the fields $X^A$ of anti de Sitter space obtained from the above formulas (\ref{ka}), when substitution is made $$\sin {\chi} \Rightarrow \sinh {\chi},\quad \cos {\chi} \Rightarrow\cosh {\chi}.$$ Note interesting in our opinion fact that in the case of $B/C<0$, equation (\ref{tah}) defines the wave equations of massive particles with mass $\mu=2\sqrt{-\frac{B}{C}}$, and if $B/C>0$ (closed model), the mass $\mu$ is imaginary and fields are tachyonic. Thus, in the latter case the gravitational constant negative and positive in the first case (if the value of $\xi$ is fixed and less than zero.)

\subsubsection{Case $Y=const$, Einstein's equations}
If we consider the presence of a perfect fluid with EMT
\be \label{mass}
 T_e = \left(
      \begin{array}{cccc}
        -\varepsilon & 0 & 0 & 0 \\
        0 & P & 0 & 0 \\
        0 & 0 & P & 0 \\
        0 & 0 & 0 & P \\
      \end{array}
    \right)
,\ee

then in the case of $Y=const=C$ holds separate law of conservation
$$\partial_{\eta}\varepsilon +3
\frac{\partial_{\eta}a}{a}(\varepsilon+P)=0,$$ and Einstein's equation is:
\be \label{ue1}
(\partial_{\eta}a)^2+a^2 k=a^4\gamma +2\varepsilon
\frac{w}{C(1+\delta_{\xi})}a^4,
\ee
 where
\be \label{ue13}
\gamma=\frac{2}{1+\delta_{\xi}}(\Lambda_X C-\frac{B}{C});
\ee

$k=\{1;-1;0\}$ respectively, for closed and open space types.

As noted above, if we "forget" about the equations (\ref{Y1}) and (\ref{gtti}), then the remaining equation (\ref{E}), in case of $Y=const$, will fully match with the equations of Einstein.
Let's proceed to the dimensionless variables
\be \label{E3}
x=\frac{t}{t_0};\quad b=b(x)=\frac{a(x)}{t_0};\quad
\tilde{C}=\frac{C}{t_0^2},\quad \lambda=\lambda(x)=\frac{\dot{b}}{b}
\ee
where $t$ -the proper time of the observer ($dt=a(\eta) \cdot d\eta$),
$t_0$ - certain scale which equals to the value of the age of the universe under consideration, dot denotes derivative by $x$. Then the equation (\ref{ue1}) takes the form:
\be \label{ue2}
\lambda^2= -\frac{k}{b^2}+\tilde{\gamma}+2\varepsilon
\frac{w}{\tilde{C}(1+\delta_{\xi})},
\ee
where $ \tilde{\gamma}=\gamma t_0^2.$

For the expected value of universe age
 $$t_0\sim13.7 mlrd.l\sim1.3 \cdot 10^{28}cm, $$ "Hubble constant"
 $H =\lambda t_0^{-1}\sim 71 km/mpk\cdot c \sim 7.7 \cdot 10^{-29}cm^{-1}$ $\Longrightarrow \lambda_0\sim 0.9963.$

There are reasons to believe that discovered by astronomers (\cite{Riess},
\cite{Perlmutter}), so-called "dark energy" - is the energy of vacuum, which is defined by the $\Lambda$ -  term \cite{Spergel}.
 We use the common value of the energy balance, where "dark energy" defined by "cosmological constant" and it is about 0.73 shares of energy.

\be \label{exper1}
 \tilde{\gamma}=0,73\lambda_0^2
\ee
Gravitational constant
\be \label{exper2}
 G_{eff} = |\frac {w} {2\xi C_0}|=6.56536^{-65}cm^2\equiv G_0,
\ee
Unknown parameters are $C=C_0, w, \Lambda_X, \xi$. Using relations (\ref{exper1}) and (\ref{exper2}) to estimate these parameters. The parameter $\xi$ can be set equal to the value of close to the conformal-invariant value $\xi_0=-1/12\Leftrightarrow \delta_{\xi}=0.$
Suppose that in the modern age the above solutions (\ref{coslam}) are carried out  (the case of the de Sitter space). Then $\gamma=B/C=H^2 \Rightarrow$  $C/B\approx 1.3048\cdot
10^{56 }cm^2,\quad |w/B| \equiv w_ 0\approx 2.855555\cdot10^{-9}cm^4.$ The distances on the order of $l_w=\sqrt[4]{w_0}\sim 0.007310cm$ correspond to parameter $w_0$. This distance is too large for the theory which leads to Planck distances ($\sim10^{-33}cm$). The emergence of such a scale is associated with a constant value $\sqrt[4]{G_0/\Lambda_{0}}\approx 0.00852 cm,$ where $G_0, \Lambda_{0}$  -modern values $G_{eff}$ and $\Lambda_{eff}$. From (\ref{coslam}) follows
$\Lambda_X \approx 2.7576\cdot10^{-113}cm^{-4}.$  Note that if we take the value $\delta_{\xi}=-3$ we get $\Lambda_X =0,$ with the same values of other parameters. Although obtained values of the constants match their experimental (modern) values , however, first of all are not suitable to define the inflationary model - too small value $H$ defines great time of inflation in solutions (\ref{masshto1}),(\ref{masshto2}). In order for our model to describe dark energy and inflation at short times it is necessary to assume that the parameters of the theory have evolved over time. Solution (\ref{coslam}) shown in the previous section is preferably used to describe the inflationary stage of the model.

Additional conditions are necessary to estimate the values of parameters. As such, we take a condition resulting from a comparison with the theory $V(\varphi)\sim\Lambda_\varphi\varphi^4/4$, where the requirement for the relative smallness of the perturbation amplitude in this theory leads to the value of $\Lambda_\varphi \sim 10^{-14}.$
Taking into account the dimensions of the parameters, this leads in our theory to the relation \be \label{exper3}
 4 \Lambda_X w \sim 10^{-14}.
 \ee
Let`s present solutions of algebraic equations (\ref{exper1}), (\ref{exper2}) and (\ref{exper3}), assuming that $\varepsilon_{vac}=1.4654 \cdot 10^{131}cm^{-4}$.
Computer calculations lead to the listed below cases.

\textbf{Case M1.} When $\delta_{\xi}=0$ and when plus sign is chosen at the disclosing of the module in the ratio (\ref{exper2}):
$
C_0=-6.23643\cdot 10^{-67}cm^{2},  w=6.82407 \cdot 10^{-132},  B=1.42485\cdot 10^{-16},
$
$
m_{bozon}= 5.96\cdot 10^{11} Gev , \Lambda_X =3.6635 \cdot10^{116}cm^{-4},
$
\be \label{exper3a}
C_{inf}=7.63804 \cdot 10^{-67}cm^{2}, H_0 =1.3658 \cdot10^{25}cm^{-1},
\ee
where $C_{inf}, \quad  H_0=\sqrt {(B/C_{inf})}$ - values $Y=C_0$ and "Hubble parameter" corresponding to the inflationary stage, calculated by the formula (\ref{coslam}), if $\Lambda_X =3.6635 \cdot10^{116} cm^{-4}$; $m_{bozon}$ - mass calculated by formula $m^2=-4\frac{B}{C_0}$ (\ref{Y1}).

 \textbf{Case M2.} When $\delta_{\xi}=0$ and when minus sign is chosen at the disclosing of the module in the ratio (\ref{exper2}), that corresponds to the choice of the minus sign in the definition (\ref{Eg4}):

$C_0=6.23643\cdot 10^{-67}cm^{2},  w=6.82 \cdot 10^{-132},$  $B=1.42485\cdot 10^{-16},$
$
m_{bozon}^2<0,  \Lambda_X =3.6635 \cdot10^{116}cm^{-4},
$
\be \label{exper3ab}
C_{inf}=7.63804\cdot 10^{-67} cm^{2}, H_0= 1.3658 \cdot10^{25}cm^{-1},
\ee

To solve the paradoxes of Friedmann cosmology (flatness, homogeneity and isotropy, etc.) it is enough for inflation to last for about 70 Hubble time. During this time, the scale factor (the initial size of about $10^{-33}cm$) increases by $e^{70} \sim 10^{33}$ times, and by the time of the Friedmann stage takes place the scale factor becomes about $10^{-3}cm $, which is required to solve the horizon problem. Thus, the stage of inflation over time $\delta t =10^{-34}c \sim  10^{-24}cm  $ "prepares" the primary matter in the size of about 0.001cm or more (\cite{linde}, \cite{luk}). For this reason "Hubble parameter" should be $H_0 \geq 10^{25}cm^{-1},$ which agrees with our model (\ref{exper3a}).

Given numerical values associated with the fine-tuning in the sense that in the expression (\ref{ue13}) for $\gamma,$ occurs numerical cancelation to the order of hundredth. To save some space we don't consider numerical values of all orders. Parameters $\Lambda_X, w $ have very weak dependance of $\xi$.
The dependence of $C_0$ on the parameter $\xi$ (in the case of {\bf M1}) is shown in Figure 1.

\begin{figure}[t]
{\includegraphics[width=6cm]{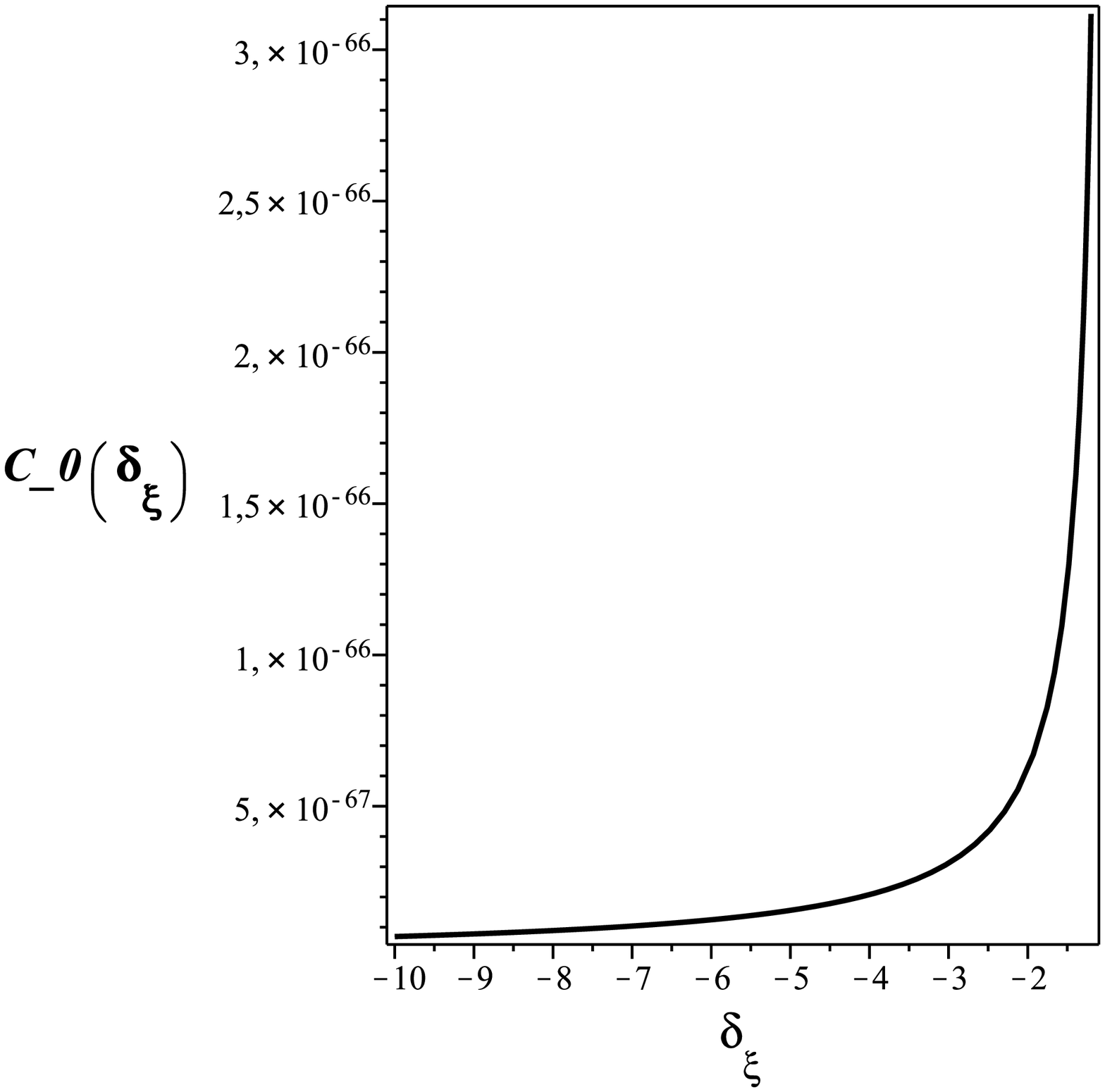}}{ \includegraphics[width=6cm]{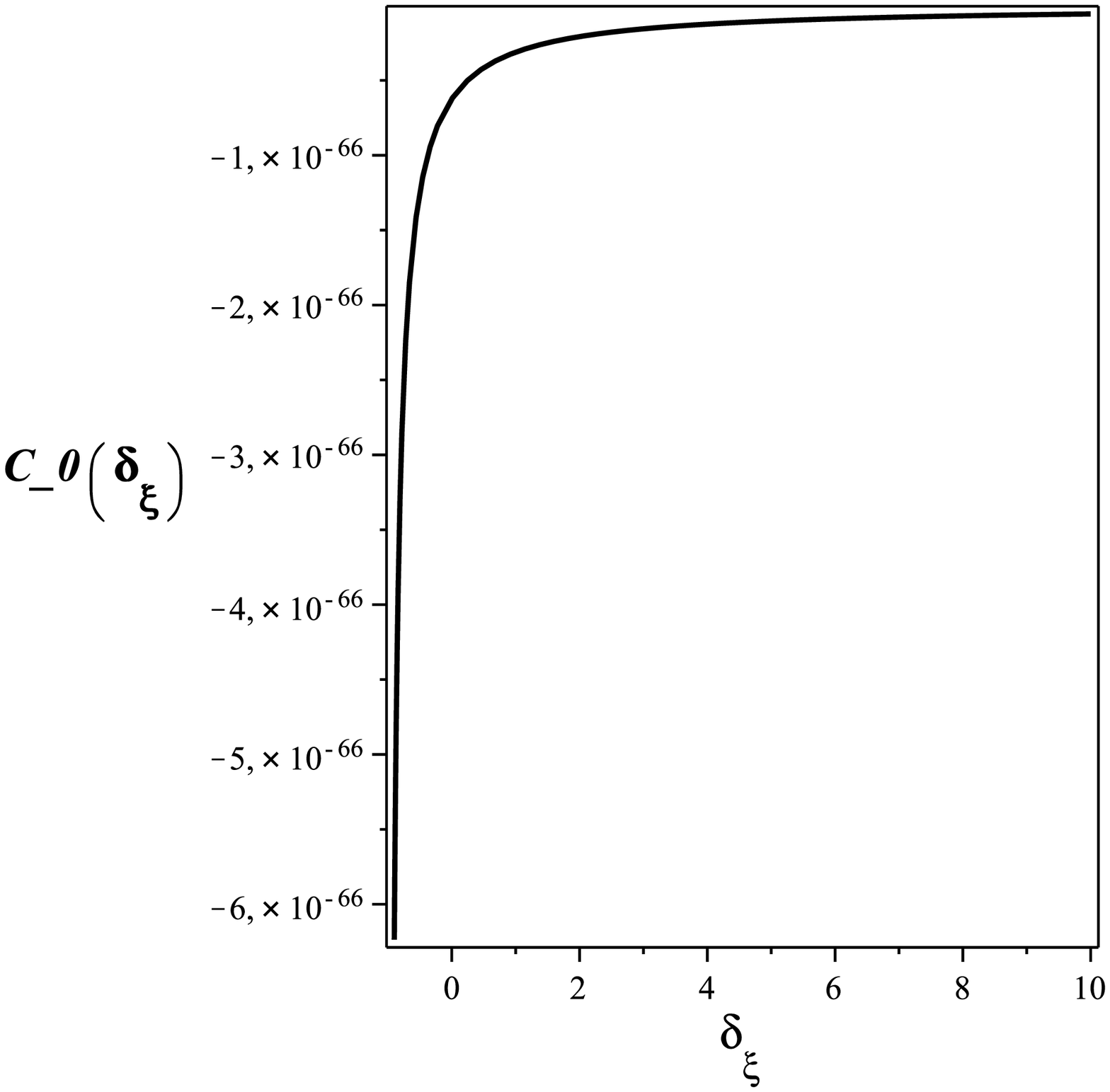}}
\caption{The dependence of $C_0$ on $\delta_\xi$: a) $\delta_\xi <-1$, \  b) $\delta_\xi >-1$
The dependence of $C_0$ on the parameter $\xi$ (in the case of {\bf M1}) is shown in Figure1.
In the case of {\bf M2} $C_0$ is multiplied by a minus. The point $\xi=0 \ (\delta_\xi=-1) $ is singular.}
\end{figure}

We draw your attention to one fact related to the original equations, corresponding to the case $S=0$. From equations (\ref{T7})-(\ref{Y4}) follows that for $S=0$, the case $Y\neq const$ is possible only for the matter in the form of a perfect fluid with equation of state
$\varepsilon-3P=-4\frac{B_0\delta_{\xi}}{\omega}$ and
  \be \label{gh1}
  \varepsilon=\frac{\varepsilon_{r0}}{a^4}-\frac{B_0\delta_{\xi}}{\omega}
  \ee
 where $\varepsilon_{r0}=const$ and $\delta_{\xi}=0$
conformal invariance is implemented.
This means that for this case in the expression (\ref{lamcos1}) we must suppose
\be \label{gh2}
\varepsilon_{vac}=-\frac{B_0  \delta_{\xi}}{w}
\ee
For this case we can determine the parameters and the value of $\xi$ as the solution of algebraic equations (\ref{exper1}), (\ref{exper2}), (\ref{exper3}) and (\ref{gh2}).
We get $$\xi=-4.3525230 \cdot10^{-7}.$$

\textbf{Case N1.} When plus sign is chosen at the disclosing of the module in the ratio (\ref{exper2}):

$C_0=-1.19402\cdot 10^{-61}cm^{2},  w=6.82403 \cdot 10^{-132},  B=5.22302\cdot 10^{-6},$
$m_{bozon}=2.61 \cdot 10^{14} Gev , \Lambda_X =3.6635 \cdot10^{116}cm^{-4},H_0= 6.61385 \cdot10^{27}cm^{-1},$
\be \label{exper3b}
C_{inf}= 1.19402\cdot 10^{-61}cm^{2}
\ee
\textbf{Case N2.}  When minus sign is chosen:
$
C_0=1.19402\cdot 10^{-61}cm^{2}, \ m_{bozon}^2<0, \ H_0= 6.61385 \cdot10^{27}cm^{-1},
$
\be \label{exper4b}
C_{inf}=1.19402 \cdot 10^{-61}cm^{2},\ee
and the rest of parameters are the same.

 Let's also consider the case when $\Lambda_X=0$ and the requirement (\ref{exper3}) is ignored. It is assumed that the description of perturbations growth mechanism uses different mechanism, not the nonlinearity in the form of $\Lambda_X Y^2$. We denote this case as $\textbf{L}$, including also the critical value $\xi=-3$.

 \textbf{Case L1.} $\delta_{\xi}=0.$ When plus sign is chosen at the disclosing of the module in the ratio (\ref{exper2}):
 $
C_0=-6.2364\cdot 10^{-67}cm^{2},  w= 6.82407 \cdot 10^{-132},  B=1.32152\cdot 10^{-123},
$
\be \label{exper5b}
m_{bozon}= 1.81 \cdot 10^{-42} Gev ,
\ee
For this case we find $C_{inf}$ based on the fact that the "Hubble parameter" corresponding to the inflationary stage

$H_0=\frac{B}{6 \xi C_{inf}}=5.99\cdot10^{25}cm^{-1} \Rightarrow $
\be \label{exper33b}
C_{inf}= -7.59281 \cdot10^{-175} cm^{2}
\ee

 Further, for the same value of $H_0$ we find.

  \textbf{Case L2.} $\delta_{\xi}=0.$ When minus sign is chosen at the disclosing of the module in the ratio (\ref{exper2}):

 $
C_0=6.23643\cdot 10^{-67}cm^{2},  w= 6.82407 \cdot 10^{-132},  B=-1.32152\cdot 10^{-123},
$
\be \label{exper6b}
m_{bozon}= 1.8 \cdot 10^{-42} Gev,  C_{inf}= 7.59281 \cdot10^{-175} cm^{2}.
\ee

  \textbf{Case L3.} $\delta_{\xi}=-3.$ When plus sign is chosen at the disclosing of the module in the ratio (\ref{exper2}):

 $
C_0=3.11821\cdot 10^{-67}cm^{2},  w= 6.82407 \cdot 10^{-132},  B=1.32152\cdot 10^{-123},
$
\be \label{exper8br}
m_{bozon}= 2.5 \cdot 10^{-42} Gev , \  C_{inf}= 3.79640 \cdot10^{-175} cm^{2}
\ee

\textbf{Case L4.} $\delta_{\xi}=-3.$ When minus sign is chosen at the disclosing of the module in the ratio (\ref{exper2}):
$
C_0=-3.11821\cdot 10^{-67}cm^{2},  w= 6.82407 \cdot 10^{-132},  B=-1.32152\cdot 10^{-123},
$
\be \label{exper9br}
m_{bozon}^2 < 0 , \  C_{inf}=- 3.79640 \cdot10^{-175} cm^{2}
\ee

 Comparing the estimated values of the field $Y$ in the modern era ($C_0$) and inflation era ($C_{inf}$) we notice that in cases of {\bf M1} and {\bf N1} are different signs but they don't differ in terms of absolute values. This means that the effective gravitational constant has different signs in these eras. The negative sign in the inflationary stage. In cases of {\bf M2} and {\bf N2} sign of $G_{eff} $ is positive and the relative change of the gravitational constant during the evolution is about 18 \% in case of {\bf M2} and 0.0001\% in case of {\bf N2}. In case of $\textbf{ L}$ value of the field $Y$ in the modern era ($C_0$) and in the era of inflation ($C_{inf}$) differ by a hundred orders of magnitude.

\section{Generalized model with the variable field. $Y\neq const$.}
Let $U =\Lambda_X (X,X)^{2} \equiv \Lambda_X Y^{2} ,$
 Let's consider the law of conservation (\ref{T72})
\be \label{u}
\partial_{\eta}Y \cdot(\xi R +2 \Lambda_X  Y)=-w[\partial_{\eta}\varepsilon +3
\frac{\partial_{\eta}a}{a}(\varepsilon+P)],
\ee
$$R=\frac{6}{a^3} (\partial_{\eta}^{2} a+ak).$$

If the energy conservation for the matter and field performed separately (left and right side of the equation (\ref{u}) equals zero), then the equations and their solutions are branched into two types (phases): 1) $Y=const$ and 2) $Y\neq const$. In order to generalize the two-phase behavior in the case of interaction of the field and matter with the energy exchange, we assume that the interaction can be decomposed into powers of $Y$.
 Let's consider the equation of state so well-known in cosmology:

1)$\varepsilon_r=3P_r$ - case of ultrarelativistic material (radiation).
 $\varepsilon_r=(\varepsilon_{r0}+Yf_{r1}+Y^2 f_{r2})/a^4$.
 The first term of this expression is the energy density of the "free" ultrarelativistic matter, and the other two terms, if such an interaction exists, describe the interaction of the matter with the field $Y$ (if not, then the coefficients $f_{r1}$, $f_{r2}$ equal to zero.)
     $\varepsilon_{r0}$,  $f_{r1}$, $f_{r2}$ - constants.

 Similarly, for the other cases of the equation of state we get :

 2) $\varepsilon_\Lambda=-P_\Lambda$ - "vacuum" state equation $\Rightarrow$
 $$\varepsilon_\Lambda=Yf_{\Lambda1} +f_{\Lambda2}Y^2 ,$$
 where $f_{\Lambda1},f_{\Lambda2}$ - constants. It is assumed that $\varepsilon_\Lambda$ contains a permanent term $\varepsilon_{vac}$, but this term was included in advance (transferred) to the parameter $B$.

3) $P=0$ - case of dust-like matter.$\Rightarrow$
 $$\varepsilon_p=(\varepsilon_{p0}+Yf_{p1}+Y^2 f_{p2})/a^3.$$

 Thus summing up for the model theory we get:

\be \label{E1}
\varepsilon=(\varepsilon_{r0}/a^4
 +\varepsilon_{p0}/a^3)+ [(Yf_{r1}+Y^2 f_{r2})/a^4 +Yf_{\Lambda1} +f_{\Lambda2}Y^2+
 (+Yf_{p1}+Y^2 f_{p2})/a^3].
 \ee

\be
 P=\varepsilon_{r0}/{3a^4}+ [(Yf_{r1}+Y^2 f_{r2})/{3a^4}- Yf_{\Lambda1} -f_{\Lambda2}Y^2].
 \ee
 Then the equation (\ref{u}) takes the form:

\be \label{Ep1}
\partial_{\eta}Y \{\xi R+2Y[\Lambda_X+w(f_{\Lambda2}+\frac{f_{p2}}{a^3}+\frac{f_{r2}}{a^4})]+
w(f_{\Lambda1}+\frac{f_{p1}}{a^3}+\frac{f_{r1}}{a^4})\}=0
\ee

Zero component of the system of equations (\ref{E}) takes the form:
\be \label{E3a}
(\partial_{\eta}a)^2+a^2 k=-\partial_{\eta}a\cdot a
\frac{\partial_{\eta}Y}{Y} +\frac{a^4}{6 \xi}(\frac{B}{Y} -\Lambda_X
Y) -\frac{w}{6 \xi Y}a^4 \varepsilon,
\ee
System (\ref{E}) is equivalent to the last two equations.

For computer modeling of solutions (\ref{Ep1}) and (\ref{E3a}) let's proceed to the dimensionless variables

$
x=t/t_0, \quad  b=b(x)=a(x)/t_0,
$
\be \label{par1}
Z=Z(x)=Y(x)/t_0^2, \quad \lambda=\lambda(x)=\dot{b}/b,
\ee
where $t$ - proper time of observer ($dt=a(\eta) \cdot d\eta$),
dot denotes the derivative with respect to $x$. Then the equations
(\ref{E3a}),(\ref{Ep1}) take form:
\be \label{Eaa4}
\lambda^2=-\frac{\dot{Z}\lambda}{Z}
-\frac{k}{b^2}-\frac{Z(\tilde{\Lambda} +w F_2)+w F_1}{6\xi}
-\frac{E w-B}{6\xi Z},
\ee
\be \label{Ex4}
\dot{Y} \{\dot{\lambda}+2 \lambda^2
+\frac{k}{b^2}+\frac{1}{3\xi}Z(\tilde{\Lambda}+wF_2)+\frac{w}{6\xi}F_1=0\}.
\ee
The last equation (taking previous one into consideration) when $\dot{Y} \neq 0$ can be rewritten as:
\be \label{Exx4}
\dot{\lambda}= \frac{k}{b^2}+\frac{1}{3\xi
Z}(w E-B)+\frac{w}{6\xi}F_1 +\frac{2\dot{Z}}{Z}\lambda.
\ee
Here we introduce the following notation:

$$\tilde{\Lambda}=\Lambda_X t_0^4+f_{\Lambda2}w, \
 F_2=\tilde{f_{p2}}/{b^3}+f_{r2}/{b^4},$$

\be \label{Egx4}
 F_1=\tilde{f_{\Lambda1}}+\tilde{f_{p1}}/{b^3}+\tilde{f_{r1}}/{b^4}, \ E=\tilde{\varepsilon_{p0}}/b^3+\tilde{\varepsilon_{r0}}/b^4;
\ee

as well as redefine the constants taking into account their dimensions

$$\tilde{\varepsilon_{p0}}=\varepsilon_{p0}/t_0^3,\quad \tilde{\varepsilon_{r0}}=\varepsilon_{r0}/t_0^4,
 \tilde{f_{\Lambda1}}=f_{\Lambda1}t_0^2,$$
$$\tilde{f_{p1}}=f_{p1}/t_0,\quad
\tilde{f_{r1}}=f_{r1}/t_0^2,\tilde{f_{p2}}=f_{p2}t_0.$$

Let's compare the equations (\ref{Eaa4}) and (\ref{Exx4}) (when $Y=const=C_0,$) with the Einstein's equations with the same EMT:

\be \label{EE4}
\lambda^2+\frac{k}{b^2}=\gamma t_0^2
+\frac{8\pi}{3}G_{eff}t_0^2\varepsilon ,
\ee

\be \label{EExx4}
\dot{\lambda}=\frac{k}{b^2}- 4\pi G_{eff}t_0^2 (\varepsilon + P),
\ee where

 $$G_{eff} = |\frac {w}{16 \pi\xi C_0 }|,  \quad
\gamma=- \frac{1}{6\xi} (-B/C_0+\Lambda_X C_0). $$ The first of these equations ((\ref{Eaa4}) and (\ref{EE4})) will match, and (\ref{Ex4}) disappear. Equation (\ref{Ex4}) equivalent to (\ref{Exx4}) (when $\dot{Y} \neq 0$) will not match with the equation (\ref{EExx4}). Recall that in the Einstein case the second equation is a differential consequence of the first.

 There are possible branching of solutions of differential equations (\ref {Eaa4}), (\ref {Ex4}) in the following ways: 1) initially (for example, in the inflationary stage) solution $Y=const \equiv C_0$ is being implemented and the scale factor is the solution of equations that coincide with Einstein equation and then, at some point of time, these solutions pass into solutions $Y \neq const $, which is the solution of equations (\ref{Eaa4}) and (\ref{Exx4}); 2) solutions are obtained in the reverse order; 3) sequential use of the first two schemes.
 There is an issue of solutions matching  $Y=const \equiv C_0$ and $Y\neq const$; as well as equation (\ref{Exx4}) may contain solutions, where $Y=const\equiv C_1$. Then occurs the problem of describing the evolution of constants - finding the solutions that convert $C_0$ into $C_1$. At the beginning it is necessary to analyze the solutions of obtained equations in simpler cases.

\subsection{Case without self-action.  $\Lambda_X =0.$}

To obtain analytical solutions, let's consider the case when in the energy density (in expansion of the interaction of field $Y$ and matter) (\ref{E1}) quadratic terms in $Y$ can be neglected. Especially since our previous estimates prove the smallness of dimensionless quantity $Z=Y/t_0^2\sim 10^{-127}$.

 The energy density corresponding to matter and interaction with the scalar field is represented in the form:
\be \label{En1}
   \varepsilon=E+Y F_1=
   \frac{\varepsilon_{r0}}{a^4}+\frac{\varepsilon_{p0}}{a^3}+Y(f_{\Lambda1}+\frac{f_{r1}}{a^4}+
   \frac{f_{p1}}{a^3}),
 \ee
 where $E, \ F_1$ corresponding functions of the scale factor $a=a(t).$
 Let's introduce the notation:
 $$\tilde{E}=-wE/(6\xi), \ \tilde{F_1}=-w F_1/(6\xi), \  \tilde{\rho}_{\Lambda}=B/(6\xi), \  \\tilde{\rho}_{r0}=-w \varepsilon_{r0}/(6\xi), \  \tilde{\rho}_{p0}=-w \varepsilon_{p0}/(6\xi),$$
\be \label{coef}
\mu_{\Lambda}=-w f_{\Lambda1}/(6\xi),\mu_{r}=-w f_{r1}/(6\xi), \  \mu_{p}=-w f_{p1}/(6\xi),
\ee

 Equation (\ref{Eaa4}) has the form:
\be \label{En3}
  \frac{\dot{a}^2}{a^2}+ \frac{k}{a^2}+\frac{\dot{a}\dot{Y} }{a Y} -\frac{1}{Y}\{ \tilde{\rho}_{\Lambda} +\tilde{E}\} - \tilde{F_1}=0,
  \ee
  where the dot - derivative of $t$.
 Equation (\ref{Ex4}) once integrated and reduced to the following form:
  \be \label{En20}
\dot{Y}\{\frac{\dot{a}^2}{a^2}+ \frac{k}{a^2}-\frac{F_0}{a^4}-\frac{2}{a^4} [\int \tilde{F_1}a^3 da]\}=0,
  \ee
 or
 \be \label{En2}
\dot{Y}\{\frac{\dot{a}^2}{a^2}+ \frac{k}{a^2}-\frac{\mu_{\Lambda}}{2}-\frac{2\mu_p}{a^3}- \frac{F_0+2\mu_r ln(a/a_0)}{a^4}+\frac{\mu_v}{a^6}\}=0,
  \ee
$F_0, \ a_0$ - integration constants.

We can prove that for the case of $\dot{Y}\neq 0$, the scale factor is the solution of equation (\ref{En20}) while $Y=Y(t)$ found by the formula:
  \be \label{En4}
  Y=\dot{a}\{c_2+\int\frac{1}{\dot{a}^3}[\tilde{\rho}_{\Lambda} +\tilde{E}]da \ \},
  \ee
  $c_2$ - integration constant.
  The proof is based on the fact that the equation (\ref{Ex4}) takes the form:
  $$\frac{\dot{a}^2}{a^2}+ \frac{k}{a^2}-\tilde{F_1}=- \frac{\ddot{a}}{a}.$$

  In such a way for the case of (\ref{En4}) solutions are in quadrature and have the form:

 1) "Einstein" (or (ES)) stage $$Y=C=const,$$
  \be \label{En5}
  \  t=\pm\int \frac{da}{\sqrt{-k +a^2(\tilde{\rho}_{\Lambda}+\tilde{E})/C +a^2 \tilde{F_1}}} \Rightarrow a\equiv a_e(t);
  \ee

  2)"restructuring" (or (RS)) stage
  \be \label{En6a}
   Y=\dot{a}\{c_2+\int\frac{1}{\dot{a}^3}[\tilde{\rho}_{\Lambda} +\tilde{E}]da \ \} \Rightarrow Y\equiv Y(t),
 \ee
  \be \label{En6}
 t=\pm \int \frac{a  da}{\sqrt{-ka^2 +F_0+2\int a^3 \tilde{F_1}da}}  \Rightarrow a\equiv a_r(t).
  \ee
 In order to resolve the integral (\ref{En6a}) it is convenient to use the formula:
  $$\dot{a}=\pm \sqrt{-k + \frac{F_0}{a^2}+\frac{2}{a^2} \int \tilde{F_1}a^3 da } .$$
 Note that solutions (\ref{En5}) and (\ref{En6}) hold for any choice of functions $\tilde{E}, \ \tilde{F_1}$,
    satisfying the generalized conservation law (\ref{u}).

From a mathematical point of view, at any given time of solution (\ref{En5}) can be transformed into solutions (\ref{En6}) and vice versa.
To describe such solutions it is necessary to match the functions of the scale factor $a(t)$ and field $Y(t)$ and their first derivatives at the point $t=t_1$ - corresponding to the moment of transition:
 \be \label{En7}
 a_e(t_1)=a_r(t_1); \ \dot{a}_e(t_1)=\dot{a}_r(t_1); \ Y(t_1)=C; \ \dot{Y}(t_1)=0.
 \ee

 These transitions are similar to the phase transition and apparently can be used to describe transition from the inflationary phase to the next phase. Transitions similar to the the first-order phase transition are described by the system, if the conditions (\ref{En7}) are supplemented by the condition of equality of second derivative at the transition point $t=t_2$:

 $a_e(t_2)=a_r(t_2), \ \dot{a}_e(t_2)=\dot{a}_r(t_2), \ \ddot{a}_e(t_2)=\ddot{a}_r(t_2),$ \
 \be \label{En8}
 Y(t_2)=Y_0, \ \dot{Y}(t_2)=0, \ \ddot{Y}(t_2)=0.
 \ee
A necessary condition for the existence of a point $t = t_2$ is the fulfillment of the following conditions on the model parameters:
 \be \label{En9}
\mu_{\Lambda}= -2 \frac{\tilde{\rho}_{\Lambda}}{Y_0} \equiv- \frac{B}{3\xi Y_0}; \ \mu_{p}=\frac{\tilde{\rho}_{p0}}{Y_0}; \ \mu_{r}=0;
 \ F_0=\frac{\tilde{\rho}_{r0}}{Y_0}.
\ee

\subsubsection{A model without interaction between the field and the matter. $F_1=0$.}

Let us assume that all the coefficients of $F_1=0$ equal to zero: $\mu_{\Lambda}=\mu_p=\mu_r=0.$ In this case there is a separate EMT matter conservation. Suppose for simplicity $\tilde{\rho}_{v0}=0$.

 Let's match the solutions (\ref{En5}) and (\ref{En6}) on the assumption that the universe was initially in the inflationary stage of expansion which is described by (ES) stage $Y=C=const$ (\ref{En5}). Following the standard approach we believe that at this stage there is no matter and the exponential expansion continues very short time. In the case of open space ($k=-1$) solution (\ref{En5}) has the form:

 $Y=C=const,$
\be \label{En10}
  a=a_e(t)=\frac {sinh \left( t{\it H_{inf}} \right) }{{\it H_{inf}}}, \   H_{inf}= \sqrt {{\frac {\tilde{\rho}_{\Lambda}}{{\it C}}}},
 \ee
This phase goes into the radiation-dominated stage, which is described by (RS) stage. Solution (\ref{En6}) has the form:
 $$
 a=a_r(t)=\sqrt {x^{2}-F_0}, \ x=t-c_1; \ Y=Y_1(x)\equiv \tilde{\rho}_{\Lambda} F_0+
$$
\be \label{En11}
+\frac {\tilde{\rho}_{\Lambda}{x}^{2}}{2 }-\frac {3 \tilde{\rho}_{\Lambda}{\it F_0}\,x \ln  \left(  \left| x+a \right| \right) }{2 \, a}+\frac {{\it c_2}\,x}{a}-\frac{\tilde{\rho}_{p_0}}{a}+\frac{\tilde{\rho}_{r0}}{F_0},
\ee
where $F_0$, $c_2$, $c_1$ -integration constants.

Substituting corresponding functions (\ref{En10})-(\ref{En11}) into algebraic equations (\ref{En7}), we find the scale factor $a_1=a(x_1)=\sqrt {x_1^{2}-F_0}$ and moment of time $x_1=t_1-c_1,$ corresponding to the transition of one solution into another.

 From equation $(-Y(x_1)+C)F_0 +\dot{Y}(x_1) \ x_1 a_1^2=0$ implies the relation:
 \be \label{E6}
 \frac{\tilde{\rho}_{p_0}}{C a_1^3}+\frac{\tilde{\rho}_{r0}}{C a_1^4}+ \frac{\tilde{\rho}_{\Lambda}}{C}=\frac{F_0}{a_1^4},
 \ee
which expresses the law of conservation of energy at the time of matching (at the time of phase transition). From this equation, we find $a_1$ and $x_1$. They are expressed via parameters $\tilde{\rho}_{\Lambda}, \ \tilde{\rho}_{r0}, \ C, F_0$. If these parameters are set, the maximum possible number of solutions for the value of the scale factor ($a_1$) equal to four, and for $x_1$ doubled. Substituting the obtained values $x_1$ into ratio $Y(x_1)=C$, we find $c_2$:
$$
c_2= - \frac {a_1}{x_1}\{\tilde{\rho}_{\Lambda} F_0+\frac {\tilde{\rho}_{\Lambda}{x_1}^{2}}{2 }-\frac {3 \tilde{\rho}_{\Lambda}{\it F_0}\,x_1 \ln  \left(  \left| x_1+a_1 \right| \right) }{2 \, a_1}-
$$
\be \label{Ec11}
-\frac{\tilde{\rho}_{p_0}}{a_1}+\frac{\tilde{\rho}_{r0}}{F_0} - C\}
\ee
Further from the first two equations of (\ref{En7}), which are dependent in this case and can be reduced to one, we find $t_1$ and $c_1$:
\be \label{Ec21}
{\it t_1}={\frac {{\it arcsinh} \left( {\it H_{inf}}\,{\it a_1}
 \right) }{{\it H_{inf}}}}, \ c_1= t_1 - x_1.
\ee
Note that we remain free to set the origin for $t$, that is, in all relationships we can make the shift $t_1\Rightarrow t_1 - t_{10}.$ The count time of inflationary stage should start with the Planck time $t_{pl}\simeq 10^{-33} cm$. So that the duration of this stage is $t_1 - t_{pl}.$
During this time the scale factor (with initial size of the order of $10^{-33}cm$) increased by $e^{70} \sim 10^{33}$ times, and by the time of the Friedmann stage becomes approximately $10^{-3}cm $ or even more, and that`s what needed to solve the horizon problem. The value of the Hubble constant at the stage of inflation is $10^{42} c^{-1} > H_{inf} > 10^{36} c^{-1}\sim 10^{26} cm^{-1}.$ Thus, the stage of inflation in less time than $\delta t \approx 10^{k_s} =10^{-34}c \sim  10^{-24}cm $ "prepares" the primary matter in the size of approximately of 0.001 cm and even more. For the numerical determination of the parameters $H_{inf}, \ F_0 $ impose a condition:
   \be \label{E7}
    a_1=\frac {sinh \left( 10^{k_s}{\it H_{inf}} \right) }{{\it H_{inf}}}= q_1.
   \ee
where $q_1, \ k_s$ are parameters for variation of initial conditions ($q_1$ approximately of 0.001 cm and more,  $-32 <k_s\leq -24$). By finding (\ref{E7}) $H_{inf}$ and using (\ref{En10}) define
 \be \label{E8}
C_{inf}=\tilde{\rho}_{\Lambda}/H_{inf}^2.
\ee

By ideology of the inflationary model inflation occurs at the time when the matter in the form of radiation and dust is not born yet. Therefore, matching of the two phases of solutions (\ref{En10})-(\ref{En11}) made at the time when matter does not exist $\tilde{\rho}_{p_0}=0,  \quad \tilde{\rho}_{r0}=0.$
Then, from relation (\ref{E6}) we find
 \be \label{E19}
F_0=a_1^4 H_{inf}^2=q_1^4 H_{inf}^2
\ee
Thus, the parameters of matching: $a_1=q_1$ - scale factor at the end of the inflationary stage which is defined by the condition (\ref {E7}), the same with $H_{inf}$; $F_0$ is found from (\ref{E19}) and $C_{inf}$ found from (\ref{E8}); $x_1=a_1^2+F_0$ is defined;  $t_1$ and $c_1$ - time of transition and shift parameter found from the formula (\ref{Ec21}) and the parameter $c_2$ defined by (\ref{Ec11}).

 By substituting the obtained constants in the formulas (\ref{En10})-(\ref{En11}), compare the values of the variables that can be obtained from our theory with the values obtained from astrophysical observations. Known observed values are: the Hubble constant - $H_0=\dot{a}/a$, the gravitational and cosmological "constants" - $\Lambda_{0} =\Lambda_{eff}(t_0)$ calculated in real time $t_0$.

 The observed values: $H_0$ - Hubble parameter, $\Lambda_{0} =3 \tilde{\rho}_{\Lambda}/Y(t_0)$ -effective cosmological constant is calculated by the formulas obtained by matching of two functions (\ref{En10})-(\ref{En11}) and independent of $\tilde{\rho}_{\Lambda}$. Indeed, if we substitute (\ref{En11}), we get:
\be \label{E9}
 Y(x)=Y_1(x)\equiv \tilde{\rho}_{\Lambda} \{F_0+\frac {{x}^{2}}{2 }-\frac {3 {\it F_0}\,x \ln  \left(  \left| x+a \right| \right) }{2 \, a}-
 \frac{x a_1}{x_1 a}[ F_0+\frac {{x_1}^{2}}{2 }-\frac {3 {\it F_0}\,x_1 \ln  \left(  \left| x_1+a_1 \right| \right) }{2 \, a_1} - \frac{1}{H_{inf}^2}]\},
  \ee
  so, from the matching and initial conditions follows that $Y(x)$ and $\tilde{\rho}_{\Lambda}$ are proportional: $ Y(t)=3 \tilde{\rho}_{\Lambda} \ \tilde{Y}(t)   \Rightarrow \Lambda_{eff}=1/\tilde{Y}.$ If the function $Y(t)$ in the modern era obeys the law of (\ref{E9}), then from (\ref{exper2}) follows that
\be \label{Ew}
w =|G_{0}  \tilde{Y} (t_{0})  \ B|  \Rightarrow G_{eff}=|\frac{G_{0 }\tilde{Y} _{0}  }{\tilde{Y}}|, \  \tilde {Y}_{0}=\tilde{Y} (t_{0}).
\ee
Thus parameters $\tilde{\rho}_{\Lambda}, \ \xi$ and  $B$ can be eliminated from the equations due to reparametrization of functions. $B=B_0-\varepsilon_{vac} w$, and we conclude that in this model,

\textbf{energy of polarization is not observed value.} At the same time the theory still has a cosmological term which has the form
\be \label{E10}
\Lambda_{eff}=1/\tilde{Y}.
\ee

The value without the influence of matter in our numerical calculations is $3.471 \times 10^{-56}cm^2$ (when $q_1=0.905, \ k_s=-24$ in formula (\ref{E7}) and $t_0=1.3 \cdot 10^{28}cm$), which is three times larger than the value adopted in the standard cosmological theory which agrees with observed data. At the same parameter values, the Hubble parameter takes $h_0=7.605 \cdot 10^{-29}cm^{-1}$, that corresponds with its present value. Thus the solution (\ref{En11}), is suitable for the explanation of the present values of the Hubble constant and possibly a cosmological constant. However, the solution of (\ref{En11}) for the scale factor does not describe the observed cosmological accelerated expansion. For our case of $F_1=0$ and the given solution: $\ddot{a}=-F_0/a^{4}<0$. (RS) stage "prepares" the values close to the observed values. Possible scenario is that at the current point of time (ES) stage is being realized, which happens as a result of the transition from the previous "restructuring" (RS). This hypothesis arises from the detailed analysis of solutions obtained using computer simulation.

Let's take into account the influence of the matter presence in the behavior of $Y(t)$. For this reason let's review the solution (\ref{En11}) with initial condition (\ref{Ec11}). We want to preserve the main stages of evolution of universe adopted in standard cosmological model and theory of hot universe (in which we assume the influence of matter on the law of expansion of the scale factor) starting from the end of the last fraction of a second of the "big bang." The initial vacuum-like state is unstable and for a split second it breaks up becoming a regular hot matter. The energy of vacuum-like state transforms to the energy of ordinary matter, the gravitational repulsion replaced by the usual gravity decelerates expansion. Recall that for small values of the scale factor it's evolution is greatly influenced by the matter in the form of radiation. Note that some fraction of radiation (photons) could be present in the stage of the "big bang", which is described in our model by (ES) stage. For inclusion of this case and the phenomenological account of radiation influence on the behavior of the scale factor (during the "big bang"), let's include "initial radiation" into the original (ES) stage. For this purpose we introduce the constant $q_r$, the value of which is small, it has no significant effect on initial inflation. Thus, we assume that radiation energy density exists in the inflationary stage
\be \label{Eiz}
\frac{\rho_{r}}{ a^4}, \ \rho_{r}=\frac {q_{r} \tilde{\rho}_{r0}}{C_{inf}},
\ee
then to describe the initial (ES) stage instead of (\ref {En10}) we get solution
$$
Y=C_{inf}=const, \quad  H_{inf}= \sqrt {{\frac {\tilde{\rho}_{\Lambda}}{{\it C_{inf}}}}},
$$
\be \label{Enr10}
a=a_e(t)=\frac {\sqrt{\sinh^2 \left( t{\it H_{inf}} \right) - H_{inf}^2  \rho_{r} \exp{(-2 H_{inf} t)}}}{{\it H_{inf}}}.
 \ee
Note that this solution for the scale factor is interesting because the function $a^2(t)$ can be continued analytically into the negative region. In this case, the metric form (\ref{met}) completely changes the sign and the scale factor converges to a constant value of $a_0^2=-1/2H_{inf}^{2}$ in finite time. Perhaps such solutions can be used to describe the preinflation stage of universe.

Solution (\ref{Enr10}) joins the solution (\ref{En6}),taking into account conditions of (\ref{En7}), which leads to the replacement of relation (\ref{E19}) by the relation:
 \be \label{En19}
 F_0=q_1^4 (H_{inf}^2)+\rho_{r},  \
\ee
 The solution describing (RS) stage (\ref{En11}) (satisfying $Y(t_1)=C_{inf}$) takes form:
  \be \label{E9a}
 Y(x)= Y_1(x)+Y_r(x)+Y_p(x),
 \ee
where
$$Y_r(x)=\frac{\tilde{\rho}_{r0}}{F_0} \ \frac{(a(x) x_1 -x a(x_1))}{a(x) x_1}, \   Y_p(x)=\tilde{\rho}_{p0} \ \frac{(x-x_1)}{a(x) x_1}.
 $$
 \be \label{E9b}
 a(x)=a_{r}(x)=\sqrt {x^{2}-F_0}, \quad x=t-c_1.
\ee
Figures 2 to 8 show graphs of a function $Y(t)$ over various time intervals, where the time is expressed in centimeters (to convert to seconds we must divide it by the speed of light). These graphs were obtained under the matching conditions:
$
a_1 \equiv a_{e}(t_{1})=a_{r}(t_{1})=0.905 cm, \
H_{inf} =5.995 \cdot 10^{25} \Rightarrow
$
\be \label{E98}
 Y(t_1)=C_{inf}=7.319 \cdot 10^{-174} cm^{2};
\ee
$
\delta_{\xi}=0,  \  t_{1}=10^{-24} cm \sim 10^{-34} sec, \ q_{r}=1.16\cdot 10^{-108},$
\be \label{E99}
  \tilde{\rho}_{r0}=15.4 \cdot 10^{-15} cm^4, \tilde{\rho}_{p0}=3.7 \cdot 10^{-38} cm^3.
\ee

\begin{figure}[t]
{\includegraphics[width=6cm]{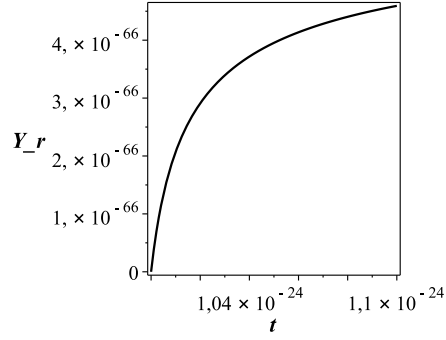}}
\caption{Relation between $Y_r$ and $t$.  $10^{-24}cm< t <1.1 \cdot 10^{-24}cm$}\label{kar2a}
\end{figure}

\begin{figure}[t]
{\includegraphics[width=6cm]{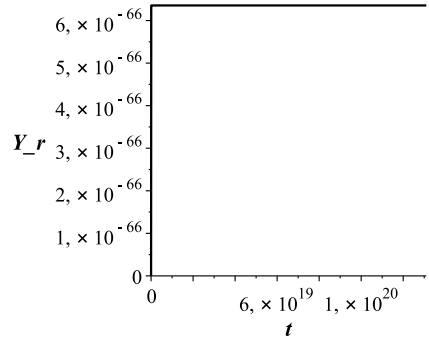}}
\caption{Relation between $Y_r$ and $t$.  $1.1 \cdot 10^{-24}cm<t<1.1 \cdot 10^{20}cm$}\label{kar2b}
\end{figure}

\begin{figure}[t]
{\includegraphics[width=6cm]{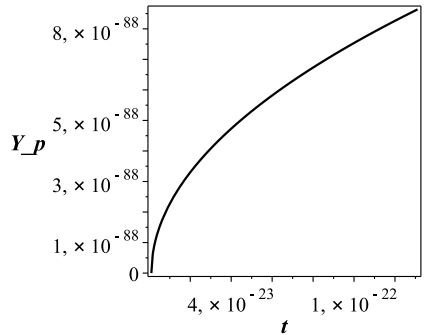}}
\caption{Relation between $Y_p$ and $t$.  $4 \cdot 10^{-23}cm< t < 10^{-22}cm$}\label{kar3a}
\end{figure}

\begin{figure}[t]
{\includegraphics[width=6cm]{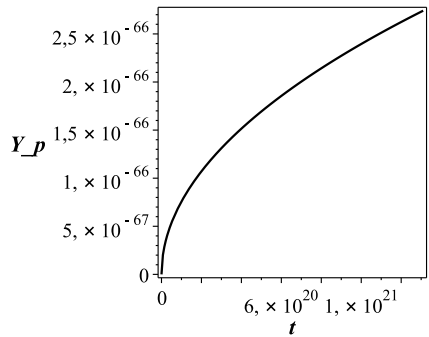}}
\caption{Relation between $Y_p$ and $t$.  $6 \cdot 10^{20}cm<t<1.1 \cdot 10^{21}cm$}\label{kar3b}
\end{figure}

\begin{figure}[t]
{\includegraphics[width=6cm]{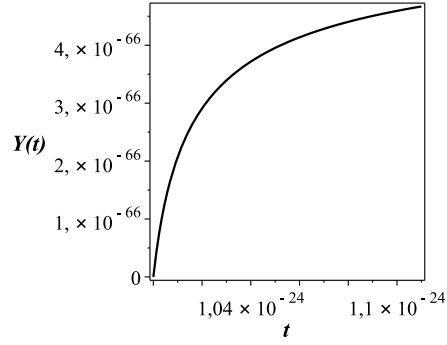}}
\caption{Relation between $Y(t)$ and $t$. In time interval $1.01 \cdot 10^{-24}< t <1.11 \cdot 10^{-24}$.} \label{kar4a}
\end{figure}

\begin{figure}[t]
{\includegraphics[width=6cm]{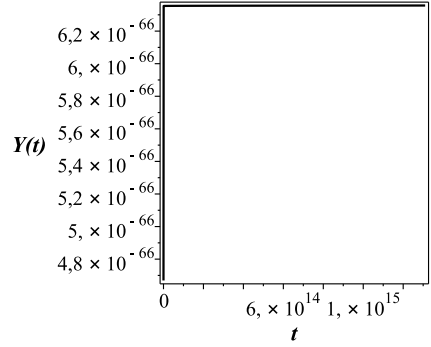}}
\caption{Relation between $Y(t)$ and $t$. In time interval $10^{-24}<t<10^{15}$ \  $Y(t)$ acts as a constant value.} \label{kar4b}
\end{figure}

\begin{figure}[t]
{\includegraphics[width=6cm]{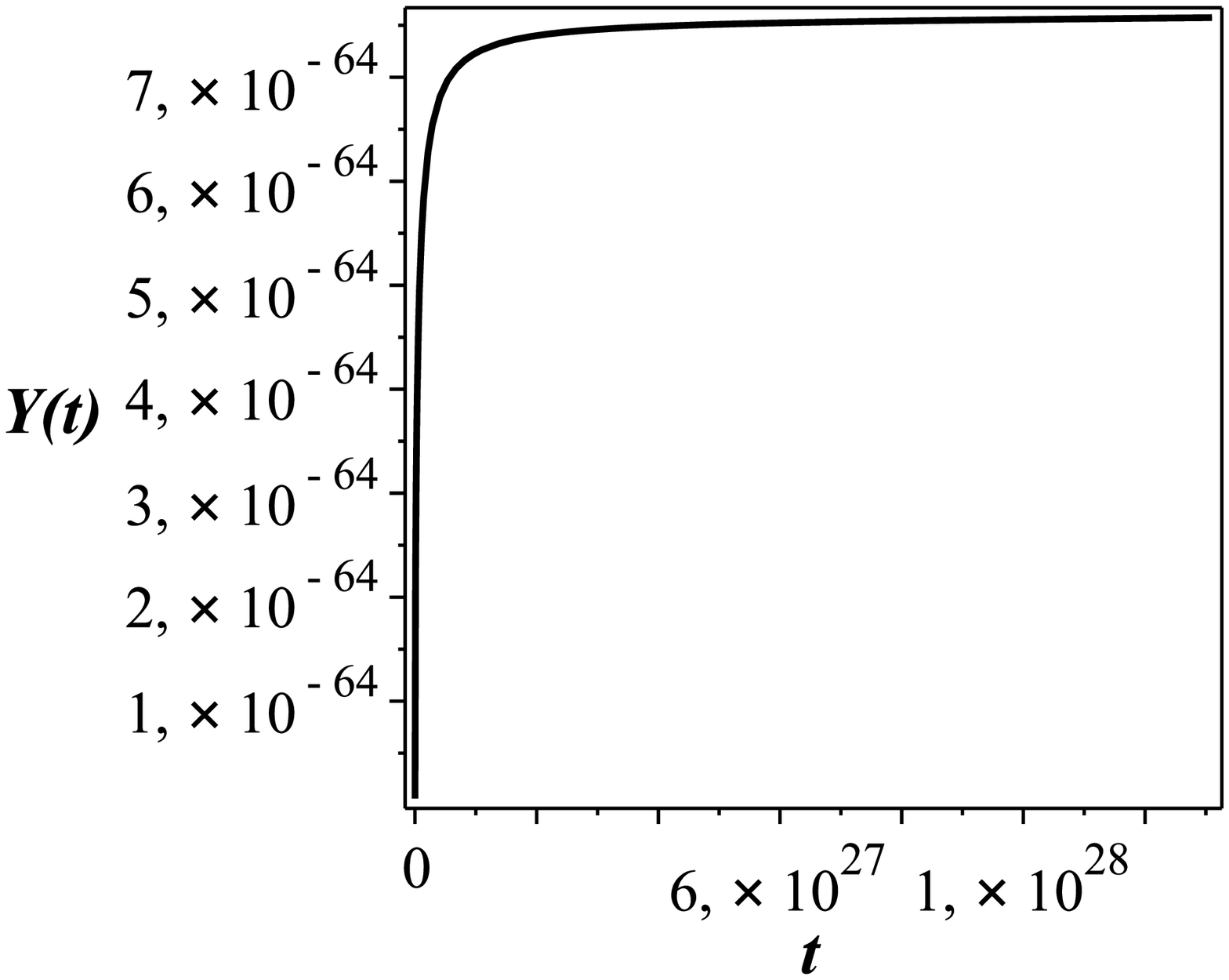}}
\caption{Relation between $Y(t)$ and $t$: in time interval $10^{15}<t<10^{28}$ function $Y(t)$ continues to grow , while transiting through the point $t_c=6.5 \cdot 10^{26}$  \ function $\ddot{Y}(t)$ changes the sign and starts a stage of slow growth of $Y(t)$.} \label{kar5}
\end{figure}

 Function $Y_{r}(t)$ in a very short time turns into the constant value (Fig. (\ref{kar2a})- (\ref{kar2b})); while function $Y_{p}(t)$ starts to affect $Y(t)$ at later times (Fig. (\ref{kar3a})-(\ref{kar3b})). Thus, the graph of $Y(t)$ (Fig. (\ref{kar4a})-(\ref{kar5})) is the sum of three function graphs: $Y_{r}(t), \  Y_{p}(t)$ and $Y_{1}(t)$.

\begin{figure}[t]
{\includegraphics[width=8cm]{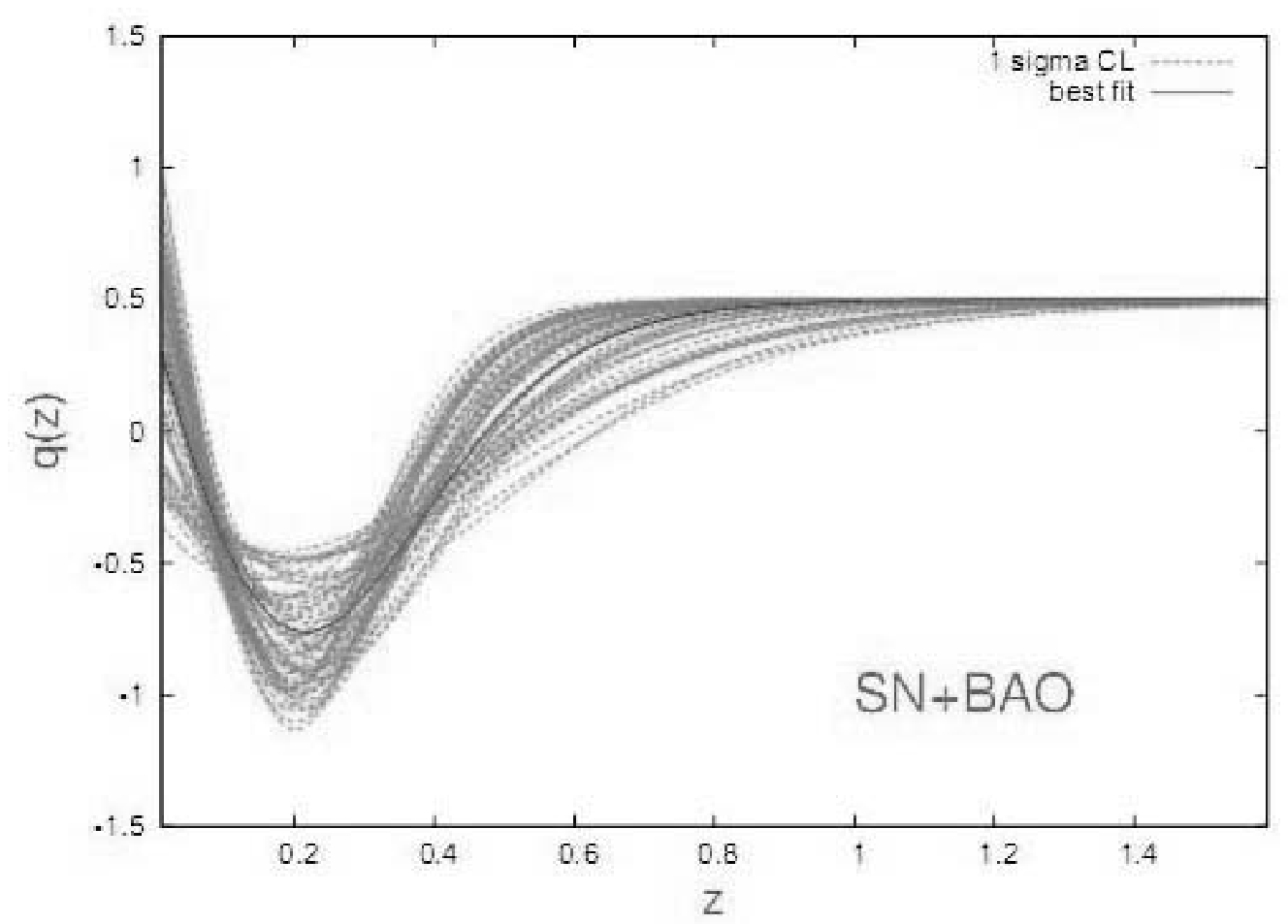}}
\caption{Recovery of the deceleration parameter by
observational data (dotted line) solid
line shows the best approximation results of
observations \cite{star}.} \label{kar6}
\end{figure}

Study of the function $Y(t)$  allows us to hypothesize that one or two times this function could switch to $Y=C_1=const$, and at these times the system was described by equations of (ES) stage. For example, this could be in the vicinity of a point $t=t_r$, when the value of $\Omega_Y (t_r)=a(t_r) \dot{Y}/(\dot{a}(t_r) Y(t_r))\ll 1.$ Then the derivative of $\dot{Y}$ in equation (\ref{E4}) for the real physical system can be neglected (for the real universe the birth of matter is not sudden but happens gradually).  For selected parameter values at the point $t_r=0.0073 cm\simeq 2.1 \cdot10^{-11} sec$ value $\Omega_Y$ takes the minimum value $\Omega_Y=2\cdot10^{-12}$.  This is to some extent agrees with the fact that in the standard cosmological model at the $10^{-10}$ second of universe life the temperature drops so that the electroweak interaction is split into weak and electromagnetic. We picked the parameters so that the value $C_1=6.32\cdot10^{-66}$ ((Fig. (\ref{kar4a})-(\ref{kar4b}))) is consistent with the expected value of the cosmological constant $\Lambda_{0}$, that is $C_1=3 \tilde{\rho}_{\Lambda}/\Lambda_{0}$ in the modern era. If this stage continues to the modern era, we will observe the values of the cosmological and gravitational constants, which follow from the astronomical and physical observations. If we choose the parameters of the theory so that the last transition into (ES) stage occurred during the time corresponding Fig. (\ref{kar4a})-(\ref{kar4b}), we obtain a theory containing at least three phase transitions and in each phase the cosmological and gravitational constants take different values. From the article (\cite{star}) that the astronomical observational data (including the construction of the light curves of SN Ia, the temperature anisotropy of the cosmic microwave background , baryon acoustic oscillations ( HLW) ) show the universe expanding acceleration which changes with the time and at times corresponding redshift $z = 0.2$ reaches its maximum value (Figure (\ref{kar6})-from  \cite{star}). This  study was not confirmed by further observational data and has not been refuted.  In our model, this time corresponds to the last phase shown in Figure 8. Note the interesting fact that the value of the field $Y$ in the last phase is about 100 times greater than its value in the previous phase (Fig. (\ref{kar4b})-(\ref{kar5})). This means that the cosmological and gravitational constants after the last transition decreased hundred times (it is necessary to take into account that due to increase of the scale factor the influence of matter on the expansion reduced dramatically). If (in the not too distant past by a cosmic scale) the value of the gravitational constant was hundred times more than in the modern era, then, according to astronomical observations associated with the effect of delay, there should be completely different physics. For example, the speed of rotation of galaxies (that is the relation of rotation speed $v(r)$ of galactic objects and a distance $r$ from the center of the galaxy) determined by the distribution of mass in the galaxy and a spherical volume of radius $r$ that encloses all the mass $m(r)$, are given by relation $v(r)=\sqrt{G_{eff} \  m(r)/r }$. Hence the conclusion that when $ z> 0.2 $, these speeds have to be ten times more than in the observations of space objects in our galaxy. Perhaps this effect can be used to explain the phenomenon of dark matter. Thus, the farther galaxy from us, the greater probability of detection of "hidden mass", which may not really exist - it is the apparent effect, which is a consequence of the evolution of the gravitational constant. It is believed that the study of motions of stars in the Milky Way did not find evidence of the presence of dark matter in a large volume of space around the Sun  \cite{Moni}.  However, we do not exclude possibility that the field $Y$ under certain physical conditions may depend on the spatial coordinates. Then there are space objects located in other phase, with other physical parameters.

\subsubsection{Model with the interaction of vacuum and matter. $F_1\neq 0$.}

 Recent observations indicate \cite{star} that the acceleration of universe expansion reached its maximum and now begins to decrease (Fig.(\ref{kar6})). This scenario is not compatible with the generally accepted standard cosmological model. In our model a scenario with the reduction of cosmological constant in the modern era (or not-distant past) is possible if we consider more complicated cases, such as the interaction expansion of field $Y$ and matter (\ref{E1}) with the quadratic terms in $Y$. However, in this case the resulting equations are complex and they can be solved only numerically. In this article we will consider only the case linear in $Y$ but considering the interaction of vacuum and matter. For this reason we assume that not all coefficients of $F_1=0$ are zero. For simplicity, let $\tilde{\rho}_{v0}=\mu_{v}=0.$ Rewrite equation (\ref{Eaa4}) fair as for ES phase (in the case of $Y=const$) as well as for $RS$ phase:
\be \label{Env3}
  \frac{\dot{a}^2}{a^2}= -\frac{k}{a^2}-\frac{\dot{a}\dot{Y} }{a  Y} +(\frac{\tilde{\rho}_{\Lambda} }{Y}+\mu_{\Lambda})+\frac{1}{a^4}(\frac{\tilde{\rho}_{r}}{Y}+\mu_{r})+ \frac{1}{a^3}(\frac{\tilde{\rho}_{p}}{Y}+\mu_{p}),
  \ee
Matching condition solutions of ES and RS stages of (\ref{E6}) is replaced by the condition:
\be \label{E66}
 \frac{\tilde{\rho}_{p_0}}{C a_1^3}+\frac{\tilde{\rho}_{r0}}{C a_1^4}+ \frac{\tilde{\rho}_{\Lambda}}{C}=\frac{F_0}{a_1^4}+ \frac{\mu_{\Lambda}}{2}+\frac{\mu_{p}}{a_1^3}+\frac{\mu_{r}}{a_1^4}(-1+ln(\frac{a_1}{a_0}))^2,
 \ee

At the critical parameter values of the theory, given by the relation (\ref{En9}), the system of equations describing the RS stage contains the solution $Y=Y_0=const$ which describes the equilibrium state of the system. The scale factor is described by the equation:
\be \label{Env4}
  \frac{\dot{a}^2}{a^2}= -\frac{k}{a^2} +\frac{1}{Y_0}(-\tilde{\rho}_{\Lambda} +\frac{\tilde{\rho}_{r}}{a^4}+ 2\frac{\tilde{\rho}_{p}}{a^3});
  \ee
which is obtained from equation (\ref{En20}) after the substitution of critical parameter values (\ref{En9}). Equation (\ref{Env4}) in the equilibrium state describes the scale factor as for RS stage as well as for ES stage. These stages can differ only by the behavior of $Y=Y(t)$ (in ES stage $Y=Y_0=const$).  In this case the integral (\ref{En6a}) is solved in the most general way, and the solution is simple:
\be \label{Yv4}
Y(t)=c_2 \dot{a} +Y_0,
\ee
where $c_2$ - constant of integration.

Lets consider a simplified model, when in the inflationary stage the matter in the form of radiation and the cosmological constant are presented and the influence of the other terms can be ignored.
\be \label{E67}
 \frac{\tilde{\rho}_{r0} q_r}{C_{inf} a_1^4}+ (\frac{\tilde{\rho}_{\Lambda}}{C_{inf}}- \frac{\mu_{\Lambda}}{2})=\frac{F_0}{a_1^4}.
 \ee
 Let the inflationary phase be described by (ES) stage ($Y=C_{inf}$) . When $\tilde{\rho}_{\Lambda}/C_{inf}\gg \mu_{\Lambda},$ scale factor (approximately) is described by the solution (\ref{Enr10}). This solution joins with the solution of (RS) stage. The solution describing (RS) stage in the analytical form is obtained for the special case of $2 \  F_0 \  \mu_{\Lambda} =1$, otherwise the solution is expressed in terms of special functions. Substituting $F_0=1/2 \mu_{\Lambda} $ in the expression (\ref{E67}) we find the value (the smallest one) $\mu_{\Lambda}$ via parameters of inflationary stage. After integration the solutions describing (RS) stage have the form:
 \be \label{E9n}
a^{2}=\frac{1}{\mu_{\Lambda}} (-1 +\exp(\sqrt{2 \mu_{\Lambda}} \ x)), \  x=t-c_{1}
 \ee
\bd
Y=c_2 \  \dot{a} +(\tilde{\rho}_{\Lambda}+\tilde{\rho}_{r0} \ \mu_{\Lambda}^{2}) \{\frac{1}{2 \mu_{\Lambda}(\mu_{\Lambda} a^2+1)}+
\frac{3(\mu_{\Lambda} a^2+1)}{4\mu_{\Lambda}^{3/2} a} \arctan(a \sqrt{ \mu_{\Lambda}})\}-
\ed
\be \label{E99n}
-\frac{\tilde{\rho}_{p0} }{2 a  (\mu_{\Lambda} a^2+1)} -\frac{5\tilde{\rho}_{\Lambda} }{4 \mu_{\Lambda}}+\frac{3\mu_{\Lambda} \tilde{\rho}_{r0}}{4},
\ee
where constant $c_2$ found from the condition $Y(t_1)=C_{inf}.$
\begin{figure}[t]
{\includegraphics[width= 6cm]{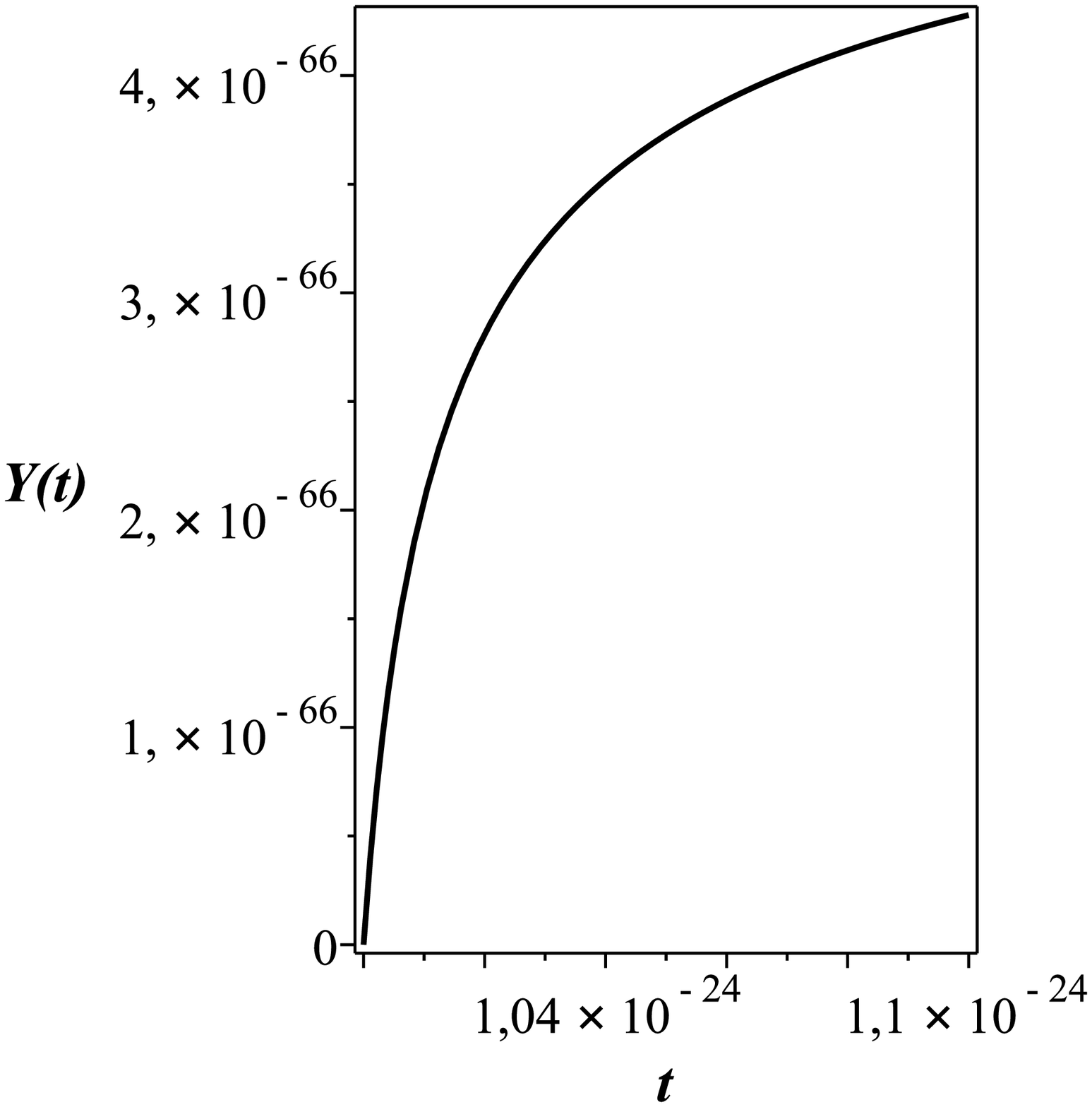}}
\caption{Relation of $Y(t)$ and $t$: when time is $10^{-24}<t<1.1 \cdot 10^{-24},$} \label{kar7a}
\end{figure}
\begin{figure}[t]
{\includegraphics[width=6cm]{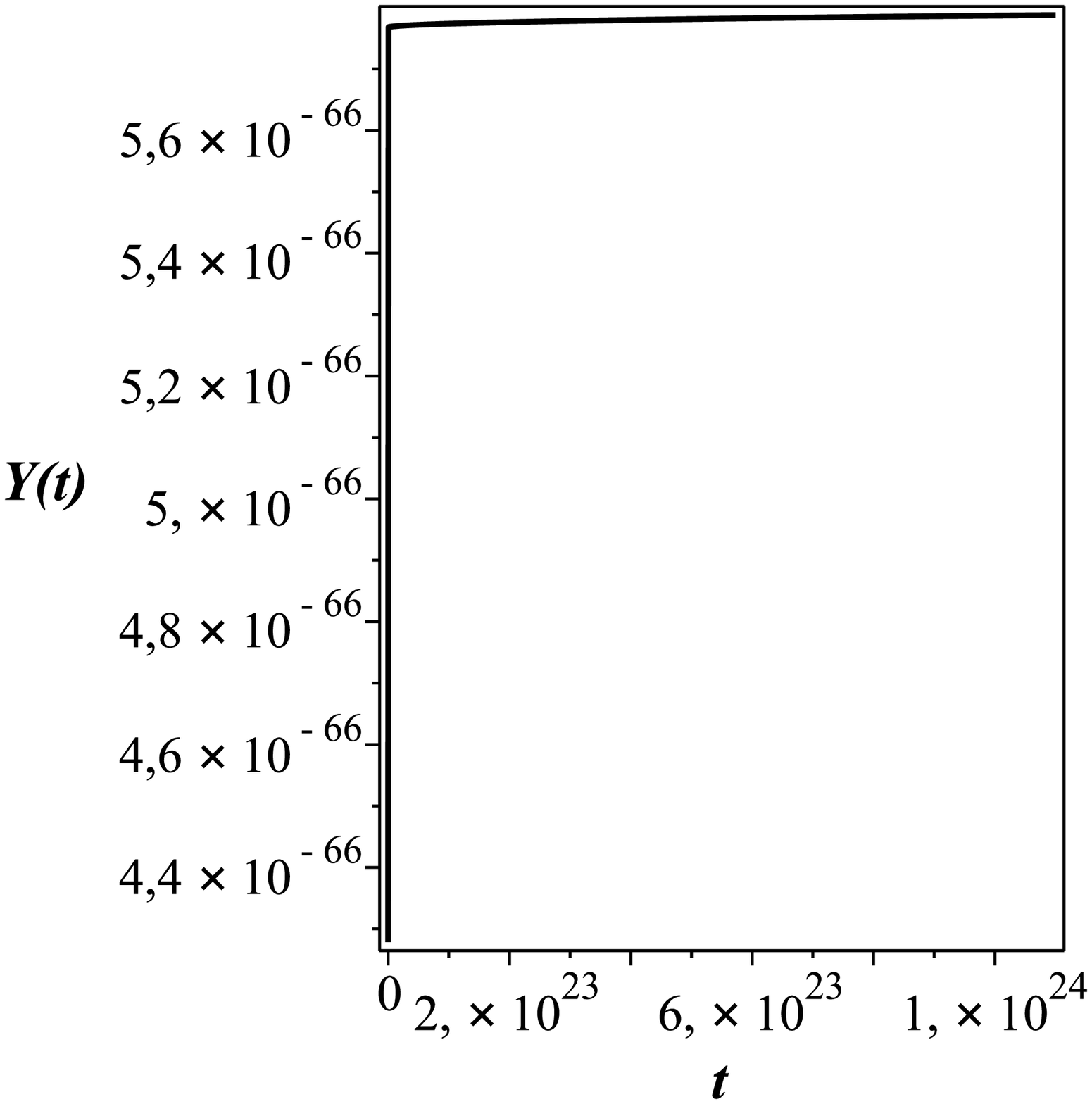}}
\caption{Relation of $Y(t)$ and $t$: when time is $ 1.1\cdot 10^{-24}<t<1.1 \cdot 10^{24}.$} \label{kar7b}
\end{figure}
\begin{figure}[t]
{\includegraphics[width=6cm]{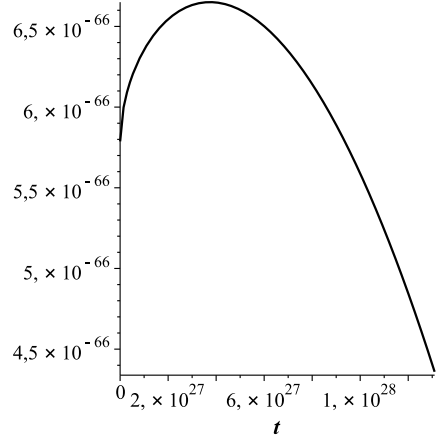}}
\caption{Relation of $Y(t)$ and $t$: when time is $1.1 \cdot 10^{24}<t<1.31 \cdot10^{28}$ the derivative of $Y(t)$ turns into zero, at the point $t_2\simeq3.7 \cdot 10^{27}cm \sim 0.26\cdot t_0$, $t_0 $- current age of the universe. } \label{kar8}
\end{figure}
Figures ((\ref{kar7a})-(\ref{kar8})) show function graphs of $Y(t)$ for different time intervals with the following parameters:
 $a_1 \equiv a_{e}(t_{1})=a_{r}(t_{1})=17967.4 cm, \ H_{inf}=6.999\cdot10^{25}cm^{-1}$
$\Rightarrow Y(t_1)=C_{inf}=-5.36 \cdot 10^{-174}cm^{2};$
$\delta_{\xi}=0, \ t_{1}=10^{-24}cm, \  q_{r}=9.3 \cdot 10^{-100},$
$\tilde{\rho}_{r0}=15.4 \cdot 10^{-15}cm^4,$
\be \label{E8r}
\tilde{\rho}_{p0}=9.25 \cdot 10^{-29} cm^3;
\mu_{\Lambda}=9.79\cdot 10^{-70} cm^{-2}.
\ee
In this model we consider the case of a negative cosmological constant $\tilde{\rho}_{\Lambda} \simeq - 2.630634 \cdot 10^{-122}. $ The initial negative value of $Y(t_1)=C_{inf}$ provides inflation of the scale factor in the inflationary stage $H_{inf}^2=\tilde{\rho}_{\Lambda} /C_{inf} >0$. The initial sign of the gravitational constant opposite to the sign which is set at the end of inflation. Suppose that by the time $t_2$ the value of $Y(t_2)$ coincides with the value that results to the modern values of the observed constants $G_{0}, \  \Lambda_{0}.$ If at the point $t_2, \ \dot{Y}(t_2)=0$ occurs transition into (ES) stage, then after this transition (as follows from formula (\ref{Env3})) the value of the cosmological constant:
 \be \label{E9lam}
\Lambda_{eff} =\frac{\tilde{\rho}_{\Lambda} }{Y(t_2)}+\mu_{\Lambda}\leq 0.
 \ee
 The observed accelerated expansion in distant galaxies is provided by a term $\mu_{\Lambda}$ that should be approximately $10^{-56}cm^2$ and this is due to the "restructuring" of (RS) stage. Thus, the behavior of the deceleration parameter (Figure 9.) for $z\rightarrow 0$ can be explained by the fact that in the recent past (on the cosmic scale) the universe has passed (ES) stage. For the more accurate calculations it is necessary to correctly simulate the behavior the scale factor, for example, abandon the condition $2 \  F_0 \  \mu_{\Lambda} =1$ and perhaps examine the influence of other terms in the equation (\ref{En2}). We can assume that in the modern era the system is close to the equilibrium state, where the solution for $Y$ takes the form (\ref{Yv4}).This decision is interesting because $\dot{Y}=c_2 \ddot{a}$, that is, transitions from $RS$ stage into $ES$ stage (and vice versa) occur at the points where the second derivative of the scale factor changes sign therefore goes from slow-motion expansion into accelerated and vice versa.
 \subsubsection{Solution to the cosmological constant problem in the theory of induced gravity.}

Astronomical observations indicate that the cosmological constant is a lot less than the value that can be obtained in the theory of elementary particles \cite{vain}. It is known in the scientific literature as the cosmological constant problem - a foothold in modern astrophysics expression meaning alleged contradiction between the predictions of two fundamental theories of physics: general relativity (GR) and quantum physics.
In the analysis of the equations of general relativity considering quantum relationships with some natural assumptions we obtain the value of the cosmological constant which is approximately the Planck-density value (\ref{evac}), while the experimental data indicates a value less by 120 orders of magnitude. Observational data on the distant type Ia supernova favor the flat model in which the universe is expanding with acceleration \cite{Riess} and to explain this phenomenon the cosmological constant value of $\Lambda_{0} =1.241 \times 10^{-56}cm^2$ is used.

In the previous sections we have considered two mechanisms of reduction of the constant part of the vacuum energy $\varepsilon_{vac}$, for the simpler case $F_1=0$. In the first case the value of $\varepsilon_{vac} $ compensated by negative density energy $-B_0/(2\xi Y) \ (B_0=1)$, the nature of which is related to the geometry - the inclusion of the manifold $\Pi$ into flat space $M$ (formula (\ref{gtti})). Reduction of two variables places requirements for the constants of theory ($w,  \ \xi, \  C_0, $). Various versions of the distribution of these parameters we reviewed earlier (\ref{exper2})- (\ref{exper9br})).

The second mechanism of reduction of the constant part of the vacuum energy is reduced to multiplicative reduction. Its principle is simple and based on the law of conservation of energy in phase transitions corresponding to different evolutionary stages of the universe and structure of the theory. Indeed, if the first phase corresponds to the stage of inflation ((ES)-stage), then the basic characteristics for physics is the Hubble constant $H_{inf}=\sqrt{3 \Lambda_{eff}} $, which is determined by the ratio of two quantities $\tilde{\rho}_{\Lambda}, \ C_{inf}$ (\ref{En10}).  Then occurs transition into (RS) stage in which the initial conditions associated with the energy conservation law at the moment of transition (\ref{E19}) define the scale factor $a(t)$ and variable field $Y(t)$ which is proportional to $\tilde{\rho}_{\Lambda}.$ From  $\Lambda_{eff}=3 \tilde{\rho}_{\Lambda}/Y(t)$ follows that  $\Lambda_{eff}$ doesn't depend of   $\tilde{\rho}_{\Lambda}=(-1+w \varepsilon_{vac})/6 \xi$,  therefore doesn't depend of $\varepsilon_{vac}$. Note that this vacuum energy reduction mechanism is similar to the mechanism of divergence reduction in quantum renormalization theory, despite the fact that this theory is classical.

Thus, the equations of this theory have solutions that can both match with the results of standard theory of gravity and can be different. This is due to the fact that the fundamental "constants" of the theory (such as the gravitational and cosmological constants) can evolve over time and also depend on the coordinates.  In the rather general case the theory describes two systems (stages): Einstein and "evolving" or "restructuring" (the name suggested by the author). Given process is similar to the phenomenon of phase transition, where the different phases (Einstein gravity system, but with different constants) pass into each other.

  An important question is: "Which of the stages corresponds to the modern era?". First, we need to look more closely at the possible correlation dependence of "dark energy" (cosmological constant), the gravitational constant and the "dark matter" on the delay parameter $z$. Second, in our model of "gravitational constant" $G_{eff}$ is inversely proportional to $Y$, for both phases. "Cosmological constant" $\Lambda_{eff}$ is constant, although different for different stages. Thus, ($RS$) is characterized by a time-varying "gravitational constant" $G_{eff}$. Suppose that in the era when $z<0.2$ (Fig. 6) in the universe is being realized a system close to the equilibrium state described by the equations (\ref{Env4}) and (\ref{Yv4}). Then
 \be \label{Yw4}
\Omega_{G}=\frac{\dot{G}_{eff}}{G_{eff}}=-\frac{\dot{Y}}{Y}=-\frac{c_2 \  \ddot{a}}{c_2 \  \dot{a}+Y_0}=\frac{c_2 \ q \  a \  H^2}{c_2 \  a H+Y_0}
\ee
 To estimate the values of (\ref{Yw4}) it is necessary to know the behavior and the value of the scale factor $a=a(t),$ Hubble functions $H=\dot{a}/a$ and deceleration parameter $q.$ Assuming $c_2 \  a H \gg Y_0,$ we get: $\Omega_{G}\sim q \ H$.
In this model the transition point from (RS) into (ES) stage is determined by the value $\dot{Y}(t_2)=0 \rightarrow \ddot{a}(t_2)=0, \ q=0.$ On the basis of Figure 6. it can be argued that the transition time $t_2$ is not too distant, and corresponds to the red-shift parameter  $z< 0.1.$ It is even possible that this time includes the time of life appearance! This means that there was a period of time when the "gravitational constant" was several times smaller than in the modern era. Thus, we need to measure the relative change in the "gravitational constant" in closely located systems of galaxies. The value of $\Omega_{G}\sim q \ H \sim q \ \cdot ( 4.5 \div
9.4)10^{-11}$ in a year.


\end{document}